\documentclass{article}

\usepackage{latexsym,amssymb}
\usepackage{graphics}
\usepackage{pstricks,psfrag}
\usepackage{amsmath}
\usepackage{subfigure}
\usepackage{vmargin}  
\usepackage{epsfig}

\newrgbcolor{myred}{1 0 0}
\newrgbcolor{mygreen}{0 1 0}
\newrgbcolor{myblue}{0 0 1}

\setmarginsrb{2.cm}{2.cm}{2.cm}{2.5cm}{0cm}{0.cm}{0cm}{1.cm}

\begin{document}

\large

\begin{titlepage}

\title{\bf Realistic Neutrino Masses from Multi-Brane Extensions
of the Randall-Sundrum Model ?}

\vskip 1cm

\author{G. Moreau \footnote{e-mail: {\tt Gregory.Moreau@ulb.ac.be}} \\ \\
{\it Service de Physique Th\'eorique, CP 225} \\
{\it Universit\'e Libre de Bruxelles, B-1050, Brussels, Belgium}}
\maketitle

\vskip 1cm

\begin{abstract} 
Scenarios based on the existence of large or warped (Randall-Sundrum model) extra dimensions
have been proposed for addressing the long standing puzzle of gauge hierarchy problem. 
Within the contexts of both those scenarios, a novel and original type of mechanism generating 
small (Dirac) neutrino masses, which relies on the presence of additional right-handed neutrinos 
that propagate in the bulk, has arisen. The main objective of the present study is to determine whether 
this geometrical mechanism can produce reasonable neutrino masses also in the interesting multi-brane 
extensions of the Randall-Sundrum model. We demonstrate that, in some multi-brane extensions, neutrino 
masses in agreement with all relevant experimental bounds can indeed be generated but at the price
of a constraint (stronger than the existing ones) on the bulk geometry, and that the other
multi-brane models even conflict with those experimental bounds. 
\end{abstract}

\vskip 1cm


\end{titlepage}

\section{Introduction}
\label{intro}

The old proposal for additional spatial dimension(s) \cite{Kaluza,Klein} and the
more recent idea of brane universe models \cite{Akama}-\cite{Antoniadis} have  
received considerable attention in the late nineties as novel frameworks for 
addressing a long
standing puzzle: the gauge hierarchy problem. Indeed, several new approaches 
toward the gauge hierarchy question, based on the existence of extra dimension(s), 
have been suggested in the literature \cite{ADD}-\cite{++B}. 
\\ The first approach \cite{ADD,AADD,ADD2}, which was proposed by Arkani-Hamed, 
Dimopoulos and Dvali (ADD), is the following one. 
If spacetime is the product of a 4-dimensional Minkowski spacetime with 
a n-dimensional compact space, and Standard Model (SM) fields are confined to 
the 4-dimensional subspace whereas gravity propagates also in the extra compact 
space, then one has,
\begin{equation}
M_{Pl}^2=M^{n+2}V_n,  
\label{ADD}
\end{equation}
$M$ being the fundamental (4+n)-dimensional mass scale of gravity, $M_{Pl} = 
1/\sqrt{8\pi G_N} \simeq 2.4 \ 10^{18}\mbox{GeV}$ ($G_N \equiv$ Newton constant) 
the effective 4-dimensional (reduced) Planck scale and $V_n$ the volume of compact 
space. Hence, by taking a sufficiently large size of new n dimensions, the value
of fundamental scale $M$ can be of order of the TeV, which removes the important 
hierarchy between the gravitational and electroweak energy scales. Nevertheless, 
another hierarchy is then introduced: the discrepancy between the electroweak 
symmetry breaking scale ($\sim 100\mbox{GeV}$) and the compactification scale 
($\sim V_n^{-1/n}$).

An elegant alternative solution to the gauge hierarchy problem was proposed 
by Randall and Sundrum (RS) \cite{Gog,RS}. The RS scenario consists of a 
5-dimensional theory in which the extra dimension (parameterized by $y$) is 
compactified on a $S^1/\mathbb{Z}_2$ orbifold of radius $R_c$ (so that $-\pi R_c \leq y 
\leq \pi R_c$). Gravity propagates in the bulk and the fifth dimension is 
bordered by two 3-branes with tensions tuned such that, 
\begin{equation}
\Lambda_{(y=0)}=-\Lambda_{(y=\pi R_c)}=-\Lambda /k=24 k M_5^3,
\label{RStensions}
\end{equation} 
where $\Lambda$ is the bulk cosmological constant, $M_5$ the fundamental 5-dimensional 
Planck scale and $1/k$ the Anti-de-Sitter ($AdS_5$) curvature radius (see below).
Within this context, the zero mode of graviton is localized on the positive tension 
brane, namely the 3-brane at $y=0$. Hence, while on this brane (referred to as 
the Planck-brane) the 4-dimensional Planck mass is of order $M_{Pl}$, on the 
other brane (at $y=\pi R_c$) the effective Planck scale, 
\begin{equation}
M_\star=wM_{Pl}, 
\label{RSratio}
\end{equation}
is affected by the exponential ``warp'' factor $w=e^{-k\pi R_c}$. From 
Eq.(\ref{RSratio}), we see that for a small extra dimension such that 
$R_c \sim 11/k$ ($k$ is typically of order $M_5\sim M_{Pl}$), one has 
$w \sim 10^{-15}$ and $M_\star={\cal O}(1)\mbox{TeV}$ (the 3-brane at 
$y=\pi R_c$ is then called the TeV-brane). If the SM particles (in particular
the Higgs boson) are confined to the TeV-brane, they feel an effective 
Planck scale $M_\star$ of the same order of magnitude as the electroweak scale. 
In this sense, the RS model provides a new natural solution to the gauge 
hierarchy problem. 
\\ Besides, in the RS framework, no additional strong hierarchy between 
fundamental scales appears (in contrast with the ADD approach) as the 
compactification scale $(2\pi R_c)^{-1}$ is of order $M_5/70$. However, 
in the RS scenario, we live on a brane with negative tension (see 
Eq.(\ref{RStensions})) which seems generically problematic as far as 
gravity and cosmology are concerned (see \cite{++A}, \cite{KaloperI}-\cite{Shiromizu} 
for a complete discussion). In particular, as it is clear from the corrected 
Friedmann equation for the Hubble expansion rate of our brane world, on our 
negative tension brane the energy density of normal matter/radiation should 
be negative, which conflicts with the absence of any noticeable effects of 
anti-gravity in our universe.

In order to avoid this potential cosmological problem, some multi-brane
extensions of the RS model (still addressing the gauge hierarchy problem), 
in which the SM fields are stuck on a positive tension brane, have been 
constructed \cite{+-+A}-\cite{++B}. 
In the RS model extension of \cite{+-+A,+-+B}, two positive 
tension branes are sitting on the fixed points of the $S^1/\mathbb{Z}_2$ orbifold, 
namely at $y=0$ and $y=\pm \pi R_c$, and a third parallel 3-brane with
negative tension can move freely in between. Within this $''+-+''$ scenario, 
our universe is the $''+''$ brane at $y=\pm \pi R_c$ (the two 
brane RS model is denoted as $''+-''$ accordingly to this terminology). 
In another possible RS extension: the $''++-''$ model \cite{+-+A,++-}, 
a $''+''$ brane is located at $y=0$, a $''-''$ brane at $y=\pm \pi R_c$ and we 
live on an intermediate parallel $''+''$ brane. In the $''++''$ model \cite{++A,++B},
two $''+''$ branes are sitting on the two orbifold fixed points and we live on 
one of them.
\\ It must be first mentioned that the original motivation for the multi-brane RS 
extensions (which is to provide a solution to the cosmological problem, arising in the 
RS model, related to the modification of Friedmann equations) is not as strong as it 
seems. Indeed, in the presence of a mechanism stabilizing the size of the extra 
dimension within the RS model (like the mechanisms suggested in \cite{GW} or \cite{LS}), 
the ordinary FRW (Friedmann-Robertson-Walker) equations are recovered \cite{Terning}.
However, the multi-brane RS extensions remain interesting alternatives to the
initial version of RS model since they give rise to a specific phenomenology. 
As a matter of fact, the multi-brane RS extensions possess a specific feature: 
the first KK excitations of a bulk fermion (or the graviton) are typically
anomalously light (relatively to the electroweak scale) \cite{Mouslopoulos,KMPR} 
\footnote{In contrast, within the pure RS framework, first KK modes of bulk 
fermions have typically electroweak-scale masses \cite{GNeubert}.}. 
The reason being that the magnitude of wave functions for first fermion KK 
modes typically approximates closely that for the 0-mode, differing significantly 
only in a region where the 0-mode wave function is exponentially suppressed.
\\ Several other models, which extend the original RS framework, have also been
elaborated in the literature, including the possibilities of configurations with
several branes \cite{OdaI,OdaII,Hatanaka}, intersecting branes \cite{Csaki,Kaloper}
or non-compact extra dimension(s) \cite{GRS,Li,RS2,LR} (as for instance in 
the case of an infinite crystalline universe \cite{++-,CryI,CryII,CryIII}). 
Those theoretical models lead generally to a rather complicated phenomenology.

Nevertheless, within all these attractive scenarios, namely ADD, RS and its 
multi-brane extensions, understanding lightness of neutrinos (with respect 
to the electroweak energy scale) becomes a challenge. Indeed, 
in this new class of brane universe models, the (effective) mass scale of 
gravity is of order of the TeV so that there exist no high energy scales. 
Hence, this brane world picture conflicts with the traditional interpretation
of small neutrino mass size invoking the `see-saw' mechanism 
\cite{seesawI}-\cite{seesawIII}, which requires a superheavy mass scale 
close to the GUT scale ($\sim 10^{16}\mbox{GeV}$). 
\\ In the context of the ADD scenario, a novel type of explanation for the 
smallness of neutrino masses has been proposed in terms of purely geometrical 
means \cite{Dudas,Russell} (the associated neutrino phenomenology has been extensively 
studied in \cite{VolFactorPhenoF}-\cite{VolFactorPhenoL}) \footnote{Let us mention that, 
within the ADD framework, other types of higher-dimensional mechanism, 
which might permit the generation of
small neutrino masses, have been suggested: the lightness of neutrinos could
result from the power-law running of Yukawa couplings \cite{Dudas} or the 
breaking of lepton number on distant branes in the bulk 
\cite{Russell,ArkaDimo}.}. 
We remind here briefly the basic ideas of this kind of
explanation, which does not rely on the existence of any high energy scale. 
The starting point is the observation that a right-handed neutrino added to
the SM would be a gauge singlet, and could thus propagate freely inside the bulk. 
In such a situation, small Dirac neutrino masses can be naturally generated as 
the Yukawa couplings, between the Higgs boson, SM left-handed neutrinos and
zero mode of bulk right-handed neutrinos, are suppressed, due to the weak
interaction probability of bulk neutrinos with SM fields (which are confined to 
our 4-dimensional subspace). This suppression is considerable since, in the ADD
framework, the volume of extra compact space is large relatively to the thickness
of the wall where SM states propagate. More precisely, the effective 4-dimensional 
mass term, between SM neutrinos and zero mode of bulk neutrinos, involves a 
suppression factor of the form (see Eq.(\ref{ADD})):    
\begin{equation}
m_\nu=\kappa \ (M^nV_n)^{-1/2}\ \upsilon=\kappa \ (M / M_{Pl})\ \upsilon, 
\label{MnuVolFac}
\end{equation}
where $\kappa$ represents the dimensionless Yukawa coupling constants and 
$\upsilon \simeq 174 \mbox{GeV}$ 
the vacuum expectation value (vev) of the SM Higgs boson. The physical 
neutrino masses are the eigenvalues of the whole neutrino mass matrix, 
which involves the masses of type (\ref{MnuVolFac}), but also the masses of the 
Kaluza-Klein (KK) excitations of bulk right-handed neutrinos, as well as the 
masses mixing those KK modes with the SM left-handed neutrinos (which originate
from the Yukawa couplings).

The same kind of intrinsically higher-dimensional mechanism as above 
(with an additional right-handed neutrino in the bulk), producing 
small neutrino masses, can apply to the RS model case
\footnote{Within the RS context, small (Dirac) neutrino masses can also
be generated by another kind of model \cite{Gherghetta}, in which the 
lepton number symmetry is explicitly broken on the Planck-brane while 
the right-handed neutrino is localized on the TeV-brane. Remarkably, 
because of the AdS/CFT correspondence, there exists a purely 4-dimensional
dual description of such models, where the right-handed neutrino 
is a composite bound state (composite right-handed neutrinos were 
independently studied in \cite{Grossman}).}. 
Nevertheless, within the RS scenario, the compactification 
volume is small ($2\pi R_c \sim 70 M_{Pl}^{-1}$) and the $AdS_5$
geometry tends to localize the 0-mode of bulk neutrino around the brane 
where are stuck SM particles (TeV-brane) 
\footnote{Unlike the 0-mode of graviton which is localized
on the positive tension branes, the 0-mode of bulk fermions are localized on 
the negative tension ones (see Eq.(\ref{RStensions})).
\label{test}}.
Hence, in order to obtain the desired order of magnitude for the effective
neutrino mass scale, via this higher-dimensional mechanism, one has to
introduce a 5-dimensional mass term for the bulk right-handed neutrino 
\cite{GNeubert} (see also \cite{seventh} for the specific case of a 
7-dimensional spacetime including one warped extra dimension). 
Indeed, such a mass term of the appropriate form can modify the wave function 
for the 0-mode\footnote{In the RS model, bulk fermions possess systematically
a 0-mode \cite{GNeubert}, in contrast with the 0-mode of bulk scalar 
\cite{RSscalarI} and vector \cite{RSvectorI,RSvectorII,RSvectorIII} 
fields which exists only for vanishing mass in the fundamental theory.} 
of bulk neutrino (applying the ideas of \cite{Rubakov,Rebbi}) 
so that its overlap with the TeV-brane, and thus its effective 4-dimensional 
Yukawa coupling with the Higgs boson and SM left-handed neutrino, appears to be 
greatly reduced.
\\ We mention here an alternative possibility, although it corresponds to a 
different physical context from the one considered in the present paper: 
in the RS model, all SM fields (except the Higgs boson, otherwise the gauge
hierarchy problem would reappear \cite{Chang,HubShafI}) could live inside the 
bulk \cite{Chang,Bajc,Pomarol}, if the SM gauge group is enhanced (in order to
satisfy the electroweak precision constraints \cite{custodial}).
This realistic hypothesis provides a new way for interpreting the flavor structure 
of SM fermion masses \cite{Pomarol,HubShafII,HubShafIII,HubShafIV}.

In the present work, we address the question which arises naturally from the
above discussion: does the same type of geometrical mechanism as above 
(with a 5-dimensional mass term for the bulk right-handed neutrino) can create 
reasonable neutrino mass values in multi-brane extensions of the RS scenario ? 
\\ One expects that indeed sufficiently reduced neutrino masses can again be
achieved through this mechanism. However, within the multi-brane RS extensions, 
the first KK excitations of right-handed neutrino can acquire ultralight masses
compared to the electroweak scale (see above).
By consequence, some mixing (induced by the Yukawa couplings) angles between
the SM left-handed neutrinos and KK excitations of right-handed neutrinos 
should be typically large. Now, one can obtain severe experimental upper bounds
on values of this kind of mixing angle between SM neutrinos and KK modes of 
right-handed neutrinos, since those KK states constitute new sterile neutrinos 
with respect to gauge interactions. In particular, these mixing angles can be 
strongly constrained by considering the experimental data on $Z^0$ boson widths 
associated to certain decay channels. 
\\ Therefore, we will derive these constraints on mixing angles (arising due to 
the presence of bulk right-handed neutrinos) which are issued from the 
measurements of $Z^0$ boson widths, within the context of 
multi-brane generalizations of the RS model. Then, we will determine whether those 
constraints translate into bounds, on the theoretical parameters, which exclude
or not multi-brane extensions of the RS model from the possible frameworks 
addressing simultaneously the gauge hierarchy and small neutrino mass questions
(the other experimental constraints on neutrino mass matrix will also be considered).

The organization of this paper is as follows. In Section \ref{AND}, 
we describe in details the entire neutrino mass matrix, within the $''+-+''$ 
framework, in case additional right-handed neutrinos propagate in the bulk. 
Then, in Section \ref{Main}, by considering this neutrino mass matrix, we
derive and discuss constraints on the $''+-+''$ scenario coming from experimental
bounds which concern neutrino physics. Based on our understanding of the $''+-+''$
analysis performed in those two sections, we discuss in Section \ref{++AND++-} 
the cases of the other multi-brane RS extensions: the $''++''$ and $''++-''$ models.
Finally, we conclude in Section \ref{conclu}.

\section{Neutrino masses in the $''+-+''$ model}
\label{AND}

\subsection{The $''+-+''$ model}
\label{Presentation}

The goal of this work is a phenomenological study on the new paradigm of anomalously 
light KK excitations which is associated to the multi-brane RS extensions. We will begin 
by concentrating on a certain concrete realization, namely the $''+-+''$ model, 
of this new paradigm because of the relative simplicity of calculations within 
the $''+-+''$ context. 
\\ We mention here that the $''+-+''$ model 
\footnote{The regions of parameter space that we consider in the present work, namely
the regions where the gauge hierarchy problem can effectively be solved, do not
correspond to the exotic Bi-gravity limit (giving rise to modifications of gravity 
at extremely large distances \cite{+-+A}) of the studied "+-+" model.}
can be seen as a limiting case of the more general $''+--+''$ brane universe scenario 
(with a flat space between the two $''-''$ branes) \cite{++-,KR}.

The $''+-+''$ model suffers from the presence of unacceptable radion fields, 
associated with the perturbations of the freely moving $''-''$ brane sandwiched between 
$''+''$ branes, which are necessarily ghost states with negative kinetic term 
\cite{RadionA,RadionB}. This fact is indeed problematic regarding the construction 
because the system is probably quantum mechanically unstable. Classically, the origin 
of the problem is connected to the violation of the weaker energy condition 
\cite{WeakEnergyA,WeakEnergyB}.

However, we will consider the $''+-+''$ model as a `toy model' and a calculation tool 
allowing us then to develop a better understanding of the phenomenological analyzes 
on the other multi-brane RS extensions, namely the $''++''$ and $''++-''$ models (which 
will be treated later: see Section \ref{++AND++-}). Now, those $''++''$ and $''++-''$ 
models represent concrete realizations of the mentioned new paradigm which do not suffer 
from the presence of radion ghost states.

Moreover, a 6-dimensional version, totally consistent from the theoretical point of view,  
of the $''+-+''$ model has been elaborated \cite{Kogan}. Within this 6-dimensional 
framework, due to the non-trivial internal space, the characteristic bounce of warp 
factor can appear even without the presence of any (moving) $''-''$ brane. Hence, one 
can avoid here the problematic presence of radion ghost states. 
\\ Now, in this 6-dimensional setup, the $''+''$ 3-branes of the $''+-+''$ model are 
replaced by 4-branes, but one of their dimensions is compact, unwarped and of Planck 
length. Thus, in the low energy limit the spacetime on the branes appears 3-dimensional. 
The 6-dimensional $''+-+''$ version has similar predictions and almost identical 
properties as the considered 5-dimensional "+-+" version, so that one can extend the 
present phenomenological study to the case of a 6-dimensional spacetime.

Besides, there exist a potential way out of the radion ghost problem arising in the 
$''+-+''$ model: the $''-''$ brane can be replaced by an external constant 4-form 
field which would mimic the effects of such a brane \cite{WayOutB,WayOutBbis}. 
\\ One can also assume that there exist a specific framework in which the radion ghost
fields, appearing in the $''+-+''$ model, condensate (so that the background can be 
stabilized) \cite{Muko} or that the theory containing ghosts may in fact be viable 
due to other kinds of particular circumstances \cite{Jeon}.
An alternative solution would be that there exist a certain regularization procedure 
(like the one proposed in the recent works \cite{regprocA,regprocB}) which applies to 
the $''+-+''$ model, resulting in a theory free from ghosts or tachyons.

\subsection{Formalism of the $''+-+''$ model}
\label{Formalism}

In view of a comparison with the $''+-+''$ scenario, we remind that
within the RS model, the considered solution to the 5-dimensional Einstein's equations,
respecting 4-dimensional Poincar\'e invariance, leads to the non-factorisable metric:
\begin{equation}
ds^2 = e^{-2 \sigma(y)} \eta_{\mu \nu} dx^\mu dx^\nu + dy^2,
\label{RSmetric}
\end{equation}
with $\sigma(y)=k |y|$, $x^\mu \ [\mu=1,\dots,4]$ the coordinates 
for the familiar 4 dimensions and $\eta_{\mu \nu}=diag(-1,1,1,1)$ the 
4-dimensional metric. The bulk geometry, associated to the metric 
(\ref{RSmetric}), is a slice of $AdS_5$ space. By considering the
fluctuations of metric (\ref{RSmetric}), one obtains, after integration over $y$,
the expression for the effective 4-dimensional Planck scale $M_{Pl}$:
\begin{equation} 
M_{Pl}^2 = \frac{M_5^3}{k} (1-e^{-2\pi k R_c}).
\label{RSkrelat}
\end{equation}

Within the $''+-+''$ model, the tensions of the $''+''$ brane sitting at $y=0$, 
the $''-''$ brane at $y=L_-$ ($L_->0$) and the $''+''$ brane at $y=L_+=\pi R_c$ 
($L_+>L_-$) are tuned to (to be compared with Eq.(\ref{RStensions})):
\begin{equation} 
\Lambda_{(y=0)}=-\Lambda_{(y=L_-)}=\Lambda_{(y=L_+)}=-\Lambda /k=24 k M_5^3. 
\label{+-+tensions}
\end{equation} 
In this framework, the metric ansatz, that should respect the 4-dimensional Poincar\'e 
invariance, is taken as in Eq.(\ref{RSmetric}) with the following solution for the 
function $\sigma(y)$,
\begin{equation} 
\sigma(y)=k(L_--||y|-L_-|). 
\label{WarpFunction}
\end{equation} 
This solution can only be trusted for an $AdS_5$ curvature smaller than the 
fundamental 5-dimensional Planck scale:
\begin{equation}
k < M_5.
\label{trust}
\end{equation}
In the $''+-+''$ extension of the RS model, the expression for the effective 
4-dimensional Planck scale $M_{Pl}$ becomes (to be compared with Eq.(\ref{RSkrelat})),
\begin{equation} 
M_{Pl}^2 = \frac{M_5^3}{k} (1-2e^{-2kL_-}+e^{-2k(2L_--L_+)}).
\label{+-+krelat}
\end{equation} 
\\ Besides, for the solution (\ref{WarpFunction}), the warp factor defined by 
Eq.(\ref{RSratio}), in which $M_\star$ denotes now the effective Planck mass on the
$''+''$ brane at $y=L_+=\pi R_c$ (where are confined all SM fields), reads as,
\begin{equation}
w=e^{-\sigma(L_+)}=e^{-k(2L_--L_+)}. 
\label{WarpFactor}
\end{equation}              
Hence, in the $''+-+''$ scenario, the gauge hierarchy problem is solved for 
$M_\star={\cal O}(\mbox{TeV})$ which is achieved when the warp factor verifies,
\begin{equation}
w \sim 10^{-15},
\label{GaugeHier}
\end{equation}
or equivalently: 
\begin{equation}
2L_--L_+\sim 34/k.
\label{GaugeHierBis}
\end{equation}
In view of future discussions, we introduce the quantity $x$ defined by,
\begin{equation}
x=k(L_+-L_-).
\label{xDEF}
\end{equation}
Then Eq.(\ref{+-+krelat}) can be rewritten in terms of the parameters $w$ 
and $x$ as: 
\begin{equation}
M_5^3=k M_{Pl}^2 [1+w^2(1-2e^{-2x})].
\label{+-+krelatWX}
\end{equation}

\subsection{Neutrino mass matrix}
\label{NeutMassMat}

In this section, we describe all the relevant contributions to the neutrino mass 
matrix. The higher-dimensional mechanism generating Dirac neutrino masses, that we 
study in this paper, requires a 5-dimensional mass term for the additional bulk 
neutrino (see Section \ref{intro}). In the $''+-+''$ background considered here, 
this mass term enters the 5-dimensional action of bulk neutrino as,  
\begin{equation}
{\cal S}_5 =\int d^4x \int dy \ \sqrt{G} \bigg ( E^M_a \bigg [ 
\frac{i}{2} \bar \Psi \gamma^a ( \overrightarrow{{\partial}_{M}}
-\overleftarrow{\partial_{M}} ) \Psi + \frac{\omega_{bcM}}{8} 
\bar \Psi \{ \gamma^a,\sigma^{bc} \} \Psi \bigg ]
-m \bar \Psi \Psi - \lambda_5 H \bar L \Psi + h.c. \bigg ), 
\label{Action}
\end{equation}
$G=det(G_{MN})=e^{-8\sigma(y)}$ (with $\sigma(y)$ as given in 
Eq.(\ref{WarpFunction})) \footnote{We use the capital indexes M, N,\dots\  
for objects defined in 5-dimensional curved space, and the lower-case indexes
a, b,\dots\ for objects defined in the tangent frame.}  
being the determinant of the metric, 
$E^M_a=diag(e^{\sigma(y)},e^{\sigma(y)},e^{\sigma(y)},e^{\sigma(y)},1)$ the 
inverse vielbein, $\Psi=\Psi(x^\mu,y)$ the neutrino spinor, 
$\gamma^a=(\gamma^\mu,i\gamma_5)$ the 4-dimensional representation of
Dirac matrices in 5-dimensional flat space, $\omega_{bcM}$ the spin 
connection ({\it c.f.} \cite{Mouslopoulos}), $m$ the neutrino mass in the 
fundamental theory, $\lambda_5$ the Yukawa coupling constant (of mass dimension 
$-1/2$), $H$ the Higgs boson field and $L$ the SM lepton doublet. 
\\ In order to localize the 0-mode of bulk neutrino, the mass $m$ must have 
a non-trivial dependence on the fifth dimension, and more precisely with a 
`(multi-)kink' profile \cite{Rubakov,Rebbi}. The mass $m$ could be the vev of 
a scalar field.
We consider the economic possibility that this scalar field has a double r\^ole,
in the sense that it also creates the branes themselves \cite{Tamvakis}
which imposes the following condition on the vev, 
\begin{equation}
m= c \ \frac{d\sigma(y)}{dy}, 
\label{VEV}
\end{equation}
where $c$ is a dimensionless parameter and $\sigma(y)$ is defined by 
Eq.(\ref{WarpFunction}). We check that the vev (\ref{VEV}) is well compatible 
with the $\mathbb{Z}_2$ symmetry ($y \to -y$) of the $S^1/\mathbb{Z}_2$ orbifold: this vev
is odd under the $\mathbb{Z}_2$ transformation (see Eq.(\ref{WarpFunction})), like 
the product $\bar \Psi \Psi$ (as fermion parity is defined by: 
$\Psi(-y)=\pm\gamma_5\Psi(y)$), so that the term $m \bar \Psi \Psi$ is even
which allows to preserve the invariance of action (\ref{Action}). 
\\ At this stage, what do we know about the value of parameter $c$ ? Since the
mass $m$ is a parameter that appears in the original 5-dimensional action  
(\ref{Action}), its natural absolute value is of order of the fundamental 
5-dimensional Planck scale $M_5$. Besides, $d\sigma(y)/dy=\pm k$ (see 
Eq.(\ref{WarpFunction})) and $k<M_5$ (see Eq.(\ref{trust})). 
Hence, it is clear from Eq.(\ref{VEV}) that the `physical' value of $c$ verifies:
$c>1$ (as discussed in \cite{Mouslopoulos}).
\\ By consequence, the relevant case is the one characterized by $c>1/2$, in which
the 0-mode of bulk neutrino is localized on the two positive tension branes 
(at $y=0$ and $y=L_+$) \cite{KMPR} like the 0-mode of graviton (as mentioned 
in foot-note \ref{test}). Therefore, the effective 4-dimensional mass $m_\nu^{(0)}$,
induced by the Yukawa coupling of action (\ref{Action}) and mixing the 0-mode of bulk 
right-handed neutrino $\psi^{(0)}_R$ with the SM left-handed neutrino $\nu_L$ (stuck 
on the $''+''$ brane at $y=L_+$), is reduced for the same geometrical reason that
the effective scale of gravity $M_\star$ (on the brane sitting at $y=L_+$) is
suppressed. This mass $m_\nu^{(0)}$ can thus be expressed (for $c>1/2$) in term of 
the warp factor $w$ defined by Eq.(\ref{RSratio}) and Eq.(\ref{WarpFactor}) 
\cite{Mouslopoulos}:
\begin{equation}
m_\nu^{(0)} \simeq \sqrt{\frac{k}{M_5}(c-\frac{1}{2})} \ w^{c-1/2} \ \upsilon. 
\label{Mnu0}
\end{equation}
In this expression, the 5-dimensional Yukawa parameter $\lambda_5$ has been taken 
to its natural value: $\lambda_5 \simeq M_5^{-1/2}$.

Similarly, the Yukawa coupling of action (\ref{Action}) also induces the following
effective 4-dimensional masses mixing $\nu_L$ with the first KK excitation of
neutrino $\psi^{(1)}_R$ \cite{Mouslopoulos}: 
\begin{equation}
m_\nu^{(1)} \simeq \sqrt{\frac{k}{M_5}(c-\frac{1}{2})} \ \upsilon, 
\label{Mnu1}
\end{equation}
or with the other KK excitations $\psi^{(n)}_R$ [$n>1$] \cite{Mouslopoulos}:
\begin{equation}
m_\nu^{(n)} \simeq \sqrt{\frac{k}{M_5}(c-\frac{1}{2})} \ e^{-x} \ \upsilon 
~~~~~~~[n=2,3,4,\dots].
\label{MnuN}
\end{equation}
The $x$ dependence in Eq.(\ref{MnuN}) can be understood as follows: the KK states
$\psi^{(n)}_R$ [$n>1$] are localized around the $''-''$ brane at $y=L_-$ (in 
contrast with the first KK mode $\psi^{(1)}_R$ which is localized on the two positive 
tension branes) so that when $x$ increases their wave function overlap with $\nu_L$ 
(trapped on the $''+''$ brane at $y=L_+$), and thus the associated mass $m_\nu^{(n)}$, 
decreases.

Finally, the excited modes of bulk neutrino $\psi^{(n)}$ [$n \geq 1$] acquire KK 
masses of the form (for $c>1/2$) \cite{KMPR}:
\begin{equation}
m_{KK}^{(1)} = \sqrt{4c^2-1} \ w \ e^{-(c+1/2)x} \ k,
\label{mKK1}
\end{equation}
\begin{equation}
m_{KK}^{(n+1)} = \xi_n \ w \ e^{-x} \ k 
~~~~~~~[n=1,2,3,\dots],
\label{mKKN}
\end{equation}
where $\xi_{2i+1}$ is the $(i+1)$-th root of $J_{c-1/2}(X)=0$ 
($i=0,1,2,\ldots$) and $\xi_{2i}$ is the $i$-th root of 
$J_{c+1/2}(X)=0$ ($i=1,2,3,\ldots$), $J_{c \pm 1/2}(X)$ denoting 
the Bessel functions of the first kind and order $c \pm 1/2$. 
We remark that the KK mass $m_{KK}^{(1)}$ of first excited state 
$\psi^{(1)}$ is manifestly singled out from the rest of the KK tower.

In conclusion, within the framework we study (namely the $''+-+''$ scenario with
an additional massive bulk neutrino), the complete neutrino mass matrix  
appears in the effective lagrangian as,
\begin{equation}
{\cal L} = - \bar \psi_L^\nu {\cal M} \psi_R^\nu + h.c., 
\label{LagMass}
\end{equation}
$\psi_{L,R}^\nu$ representing the 4-dimensional fields: 
$\psi_L^\nu=(\nu_L,\psi^{(1)}_L,\psi^{(2)}_L,\ldots)$ and 
$\psi_R^\nu=(\psi^{(0)}_R,\psi^{(1)}_R,\psi^{(2)}_R,\ldots)$, and reads as 
(see Eq.(\ref{Mnu0})-(\ref{mKKN})),
\begin{equation}
{\cal M} = 
\left( \begin{array}{cccc}
m_\nu^{(0)} & m_\nu^{(1)} & m_\nu^{(2)} & \ldots \\
0 & m_{KK}^{(1)} &  0 &  \ldots \\
0 & 0 & m_{KK}^{(2)}  & \ldots  \\
\vdots & \vdots &  \vdots &  \ddots 
\end{array} \right).
\label{MassMatrix}
\end{equation}

\section{Experimental constraints on the $''+-+''$ model from neutrino physics}
\label{Main}

\subsection{Number of neutrino generations}
\label{Ngen}

\subsubsection{Bound from $Z^0$ width measurements}
\label{Measure}

If the SM left-handed neutrino mixes with some sterile (with respect to 
the SM gauge interactions) left-handed neutrinos, then its effective weak
charge is diminished \cite{HubShafIII}. That is the reason why, in such a 
situation, the measurements of $Z^0$ boson width induce a constraint on mixing 
angles between sterile left-handed neutrinos and the SM one. In our framework, 
we must study this constraint since the SM neutrino $\nu_L$ mixes with the 
KK excitations of bulk neutrino $\psi^{(n)}_L$ [$n \geq 1$] (see Section 
\ref{NeutMassMat}), which constitute sterile left-handed neutrinos.
\\ Here, we derive this constraint in the considered physical context. Let us
define the quantity $n_\nu$ as,
\begin{equation}
n_\nu = \frac{\Gamma^{exp}(Z^0 \to \mbox{invisible})}
{\Gamma^{th}(Z^0 \to \bar\nu_L\nu_L)}, 
\label{nRatio}
\end{equation}
where $\Gamma^{exp}(Z^0 \to \mbox{invisible})$ stands for the experimental
data on $Z^0$ boson width associated to the decay channel into any undetectable 
particle, and, $\Gamma^{th}(Z^0 \to \bar\nu_L\nu_L)$ represents the known 
theoretical prediction of $Z^0$ boson width associated to the decay into a single
family of SM neutrino (neutrino masses being neglected relatively to the $Z^0$ 
mass). The value of $n_\nu$ obtained experimentally is \cite{Sirlin},
\begin{equation}
n_\nu = 2.985 \pm 0.008  
\label{nBound}
\end{equation}
In the absence of any sterile neutrino effect, $n_\nu$ is nothing but an 
experimental estimation of the number of SM neutrino generations. 
Since in our framework, the SM 
neutrinos mix not only with each other but also with the sterile neutrinos 
$\psi^{(n)}_L$, their effective weak coupling is suppressed so that the number of 
SM neutrino generations reads as \cite{GNeubert},
\begin{equation}
n_\nu^{gen} = 3 = n_\nu/\cos^2\theta_\nu, 
\label{nReal}
\end{equation}
where $\cos^2\theta_\nu$ represents the admixture of lightest neutrino
eigenstates for the SM electron neutrino, in case this admixture is identical for
the muon and tau neutrinos. The lightest neutrino eigenstates stand here for all the
neutrino eigenstates with a mass smaller than the $Z^0$ boson mass (so that they can 
be produced in the $Z^0$ decay). The quantity $\cos^2\theta_\nu$ involves thus
mixing angles between SM left-handed neutrinos and sterile neutrinos $\psi^{(n)}_L$.
Eq.(\ref{nBound}) and Eq.(\ref{nReal}) lead to the (expected) bound on
this angle $\theta_\nu$: 
\begin{equation}
\tan^2\theta_\nu<0.0077 
\label{thetaBound}
\end{equation}
What is the precise definition of $\cos\theta_\nu$ in the case of a
unique lepton flavor (the three flavor case will be treated in Section 
\ref{3Flavors}) ? This definition is,
\begin{equation}
\cos\theta_\nu=U_{01}, 
\label{DEF}
\end{equation}
where $U$ is the unitary matrix responsible for the basis transformation:
\begin{equation}
\psi^{\nu}_L=U \ \psi^{phys}_L, 
\label{UmatDEF}
\end{equation}
$\psi^{phys}_L$ containing the neutrino
mass eigenstates: $\psi^{phys}_L=(\nu_1,\nu_2,\nu_3,\ldots)$ (and
$\psi^{\nu}_L$ being defined in Section (\ref{NeutMassMat})). The two
indexes $0$ and $1$ of matrix element (\ref{DEF}) correspond respectively to the 
SM neutrino $\nu_L$ in vector $\psi^{\nu}_L$ and to the lightest neutrino 
eigenstate $\nu_1$ in vector $\psi^{phys}_L$. More explicitly, one has
$\nu_L=U_{01}\ \nu_1+\ldots$ Here, we have assumed that only the lightest 
neutrino eigenstate $\nu_1$ has a mass smaller than the $Z^0$ mass (The hypothesis
that more neutrino eigenstates are lighter than the $Z^0$ boson will be discussed
in Section \ref{PARAMx}).

\subsubsection{Implications for the parameters of the $''+-+''$ model}
\label{implication}

In this part, we will translate the experimental bound (\ref{thetaBound}) into
a constraint on the fundamental parameters in the version of the $''+-+''$ scenario
with a massive bulk right-handed neutrino.
\\ The mixing angle $\theta_\nu$ entering Eq.(\ref{thetaBound}), and defined in 
Eq.(\ref{DEF})-(\ref{UmatDEF}), is calculated in Appendix \ref{ApMix},
in case the neutrino mass matrix is given by Eq.(\ref{MassMatrix}) which is 
characteristic of the presence of a bulk right-handed neutrino (eigenvalues of 
the hermitian square of matrix (\ref{MassMatrix}) are discussed in Appendix 
\ref{ApEigen}). The result appears in Eq.(\ref{wanted}). The KK masses and masses 
mixing the SM neutrino with excited states of bulk neutrino, which enter 
Eq.(\ref{wanted}), can be replaced by their expression within the $''+-+''$ 
framework given in Eq.(\ref{mKK1})-(\ref{mKKN}) and Eq.(\ref{Mnu1})-(\ref{MnuN}) 
respectively: this leads to the following expression for $\tan^2\theta_\nu$ in terms 
of the fundamental parameters, 
\begin{equation}
\tan^2\theta_\nu \simeq 
\frac{\upsilon^2}{w^2kM_5}
\bigg [ \frac{e^{(2c+1)x}}{2(2c+1)}
+ (c-\frac{1}{2}) \bigg ( 
\sum_{i=1}^\infty \frac{1}{(\zeta^+_i)^2} +
\sum_{i=1}^\infty \frac{1}{(\zeta^-_i)^2}
\bigg ) \bigg ], 
\label{thetaA}
\end{equation}
where $\zeta^+_i$ and $\zeta^-_i$ are the $i$-th roots ($i=1,2,3,\ldots$) of 
$J_{c+1/2}(X)=0$ and $J_{c-1/2}(X)=0$ respectively. The two infinite sums of 
Eq.(\ref{thetaA}) can be performed exactly and yield, 
\begin{equation}
\tan^2\theta_\nu \simeq 
\frac{\upsilon^2}{w^2kM_5}
\bigg [ \frac{e^{(2c+1)x}}{2(2c+1)}
+ g(c) \bigg ], \ \ \mbox{with} \
g(c)=\frac{(c+1)(2c-1)}{(2c+3)(2c+1)}.  
\label{thetaB}
\end{equation}
From this expression of $\tan^2\theta_\nu$ and the experimental bound 
(\ref{thetaBound}) on $\tan^2\theta_\nu$, we deduce the following constraint
on the theoretical parameters $x,w,k,M_5$ and $c$ (to be added to the SM parameters) 
of the $''+-+''$ model with an additional massive bulk neutrino,
\begin{equation}
x \lesssim \frac{1}{2c+1}
\ln \bigg [ 
0.0077 \times 2(2c+1) \frac{w^2kM_5}{\upsilon^2}
- \frac{(2c-1)(2c+2)}{(2c+3)}
\bigg ].
\label{FINAL}
\end{equation}
The $\ln$ function involved in Eq.(\ref{FINAL}) is well defined on the intervals 
of parameters that we will consider (see Eq.(\ref{lnOK})).
\\ It is instructive to remark that, in fact, the upper bound (\ref{thetaBound}) 
on $\tan^2\theta_\nu$ has been expressed as an upper bound on the parameter 
$x$ ({\it c.f.} Eq.(\ref{FINAL})). This point can be understood in a physical 
way as follows. The dominant effect of a decrease of $x$ on the matrix (\ref{MassMatrix})
is that the KK masses (\ref{mKK1})-(\ref{mKKN}) for excited modes of bulk neutrino 
increase, so that these excited modes $\psi^{(n)}$ [$n \geq 1$] tend to decouple.
This induces a decrease of the mixing, quantified typically by $\tan^2\theta_\nu$
(see Section \ref{Measure}), between the left-handed component $\psi^{(n)}_L$ of those
sterile neutrinos $\psi^{(n)}$ [$n \geq 1$] and the SM neutrino $\nu_L$.

\subsection{Neutrino masses}
\label{NeutMass}

In this section, we will determine the constraints on parameters of the $''+-+''$ scenario, 
with an additional massive bulk neutrino, originating from the experimental bounds on neutrino 
masses. We will concentrate on the experimental bounds on absolute neutrino mass scales: here, 
the relevant bounds are those extracted from the tritium beta decay experiments 
\cite{Review,Mainz,Troitsk,Katrin} since those bounds are independent of whether neutrinos 
are Majorana or Dirac particles. In contrast, the other bounds issued from neutrinoless double 
beta decay experiments (see \cite{DoubleBeta} for a review) apply only on Majorana neutrino 
masses and thus do not hold in the present framework where neutrinos acquire Dirac masses (see
Eq.(\ref{LagMass})). 
\\ The best limit coming from data on tritium beta decay has been obtained by the Mainz 
experiment and reads as \cite{Mainz},
\begin{equation}
m_\beta \leq 2.2 \ \mbox{eV} \ \ \ (95\% \ C.L.). 
\label{TritiumBound}
\end{equation}
We also indicate the limit extracted from data on tritium beta decay measured by the Troitsk 
experiment: $m_\beta \leq 2.5 \ \mbox{eV} \ (95\% \ C.L.)$ \cite{Troitsk}. 
In the assumption of no mixing between lepton flavors, the quantity $m_\beta=m(\nu_e)$ 
introduced in Eq.(\ref{TritiumBound}) is the electron neutrino mass, or equivalently the 
associated neutrino mass eigenvalue \cite{Review,Katrin}. Therefore, in our physical
context with only one lepton flavor (the electron flavor), the experimental limit 
(\ref{TritiumBound}) can be applied on the smallest neutrino mass eigenvalue $m_{\nu_1}$ 
(see end of Section \ref{Measure} and Eq.(\ref{DIAGO})) which leads then to, 
\begin{equation}
m_{\nu_1} \leq 2.2 \ \mbox{eV}. 
\label{TritiumApplied}
\end{equation}
It must be mentioned that under the realistic hypothesis of three mixing lepton flavors, 
the effective mass, to which are sensitive the tritium beta decay experiments, reads as 
$m_\beta=( \sum_{i=1}^{3} \vert U^{MNS}_{ei} \vert ^2 \ m_{\nu_i}^2 )^{1/2}$ where $U^{MNS}$
is the lepton mixing matrix \cite{Review,Katrin}.
In this case, taking into account the experimental lepton mixing angle values \cite{SquaDiff}
and small squared neutrino mass differences 
($\vert m_{\nu_2}^2-m_{\nu_1}^2 \vert \in [6.1,8.4] \times 10^{-5} \mbox{eV}^2$ and 
 $\vert m_{\nu_3}^2-m_{\nu_1}^2 \vert \in [1.4,3.0] \times 10^{-3} \mbox{eV}^2$ at $2 \sigma$ from a 
global data analysis \cite{SquaDiff}
\footnote{The study performed in \cite{SquaDiff} is based on the results of the
atmospheric and solar neutrino experiments as well as the accelerator (K2K)
and reactor (CHOOZ and KamLAND) experiments.}) 
obtained from oscillation experiments, one would expect that 
Eq.(\ref{TritiumBound}) still leads to upper bounds on the three weakest neutrino mass eigenvalues 
$m_{\nu_{1,2,3}}$ of order of the $\mbox{eV}$ (as in Eq.(\ref{TritiumApplied})). 
\\ Besides, in the case of three lepton flavors, the upper cosmological bound 
$\sum_{i=1}^{3} m_{\nu_i} <0.7 \leftrightarrow 1.01 \mbox{eV}$ 
(depending on cosmological priors), which comes from 
WMAP and 2dFGRS galaxy survey \cite{Hannestad}, also corresponds to limits on the three weakest 
neutrino mass eigenvalues $m_{\nu_{1,2,3}}$ of the same order as in Eq.(\ref{TritiumApplied}).

An expression for the smallest neutrino mass eigenvalue $m_{\nu_1}$, which enters 
Eq.(\ref{TritiumApplied}), can be found by combining Eq.(\ref{ResultI}) and Eq.(\ref{wanted}).
The result is, 
\begin{equation}
m_{\nu_1} \simeq \cos \theta_\nu \ m_\nu^{(0)}.
\label{combine}
\end{equation}
The expression (\ref{combine}), together with Eq.(\ref{Mnu0}) and Eq.(\ref{TritiumApplied}), 
leads to the following experimental constraint on fundamental parameters,
\begin{equation}
\cos \theta_\nu \ \sqrt{\frac{k}{M_5}(c-\frac{1}{2})} \ w^{c-1/2} \ \upsilon
\lesssim 2.2 \ \mbox{eV}. 
\label{FINALbis}
\end{equation}

\subsection{Combination of the constraints}

\subsubsection{Constraint on the parameter $c$}
\label{PARAMc}

Here, we deduce from Eq.(\ref{FINALbis}) (representing an experimental bound on
neutrino mass) a constraint on theoretical parameter $c$ (defined by Eq.(\ref{VEV})).  
For the value of ratio $k/M_5$ involved in Eq.(\ref{FINALbis}), we consider the range
$10^{-4} \lesssim k/M_5 \lesssim 1$, its upper boundary being motivated by Eq.(\ref{trust})
and its lower one by the fact that it is not desirable to introduce a new high hierarchy
between the $AdS_5$ curvature $k$ and the fundamental scale of gravity $M_5$.
Hence, by using the value of warp factor given in Eq.(\ref{GaugeHier}) (for which the
gauge hierarchy problem is solved), the bound (\ref{thetaBound}) and
Eq.(\ref{FINALbis}), we obtain the numerical results:  
\begin{equation}
c \gtrsim 1.08 \ \ \ \mbox{for} \ k/M_5=10^{-4},   
\label{climA}
\end{equation}
\begin{equation}
c \gtrsim 1.22 \ \ \ \mbox{for} \ k/M_5=1.   
\label{climB}
\end{equation}
Those results mean that the obtained value of the lower limit on parameter $c$ 
lies typically in the interval $[1.08,1.22]$ if $10^{-4} \lesssim k/M_5 \lesssim 1$. 
The dependence of this lower limit for $c$ on the warp factor value is weak: 
for instance, if $w=\{10^{-13};10^{-14};10^{-15};10^{-16};10^{-17}\}$ then 
Eq.(\ref{climB}) reads as $c \gtrsim \{1.33;1.27;1.22;1.17;1.13\}$ [$k/M_5=1$] 
respectively.

\subsubsection{Constraint on the parameter $k$}
\label{PARAMk}

In fact, the bound (\ref{FINAL}) on parameter $x$ allows to impose a
constraint on the $AdS_5$ curvature $k$. Let us derive this constraint. For
that purpose, we first observe that, within the considered $''+-+''$ model,
the quantity $x$ defined by Eq.(\ref{xDEF}) is positive (see Section
\ref{Formalism}), namely, 
\begin{equation}
x>0.   
\label{xPOS}
\end{equation}
Indeed, the opposite case $x<0$ ($\Leftrightarrow 0<L_+<L_-$) corresponds to 
the brane configuration of the $''++-''$ scenario, in which a $''+''$ brane 
(at $y=L_+$) is sitting between an other $''+''$ brane (at $y=0$) and a 
$''-''$ brane (at $y=L_-$). Eq.(\ref{FINAL}) and Eq.(\ref{xPOS}) lead to, 
\begin{equation} 
0.0077 \times 2(2c+1) \frac{w^2kM_5}{\upsilon^2}
- \frac{(2c-1)(2c+2)}{(2c+3)} \gtrsim 1.
\label{lnOK}
\end{equation}
By considering Eq.(\ref{lnOK}) together with the limit (\ref{climA}), 
which is conservative all through the considered range $10^{-4} \lesssim k/M_5
\lesssim 1$ (see Section \ref{PARAMc}), we find the following numerical
result, 
\begin{equation}
w\sqrt{kM_5} \gtrsim 1097 \ \mbox{GeV}.   
\label{wkM5}
\end{equation}
Now, by taking into account the relevant values of brane positions
$L_-$ and $L_+$ (see Eq.(\ref{L1L2}) that is presented later), the 
relation (\ref{+-+krelat}) (characteristic of the $''+-+''$ model) can 
be rewritten in a good approximation as,
\begin{equation} 
M_5^3 \simeq k M_{Pl}^2.
\label{M5exp}
\end{equation} 
This new relation, the condition (\ref{trust}) and the value of warp 
factor (\ref{GaugeHier}) (required in order to solve the gauge hierarchy 
problem) can be combined to give $w\sqrt{kM_5} \leq {\cal O}(\mbox{TeV})$, 
which leads, together with Eq.(\ref{wkM5}), to the important result:
\begin{equation}
w\sqrt{kM_5} = {\cal O}(\mbox{TeV}).   
\label{wkM5bis}
\end{equation}
This result and the $M_5$ expression (\ref{M5exp}) give rise to the following 
expected constraint on curvature parameter $k$ (for the value of $w$ given by
Eq.(\ref{GaugeHier})):
\begin{equation}
k \sim M_5 \sim M_{Pl}.   
\label{Kbound}
\end{equation}

Let us make an important comment on the relation found in Eq.(\ref{Kbound}): 
this relation $k \sim M_5$ can be interpreted as a condition of `naturality'
(fixing the $AdS_5$ curvature parameter $k$) within the generic $''+-+''$ 
framework. In the sense that this relation avoids the possibility to 
introduce a new strong hierarchy between the energy scale $k$ and 
fundamental Planck scale $M_5$ in the $''+-+''$ model (the main interest 
of the $''+-+''$ model being to solve the problem of strong hierarchy 
between the electroweak scale and scale $M_5$).

\subsubsection{Constraint on the parameter $x$}
\label{PARAMx}

\begin{figure}[!t]
\begin{center} 
\psfrag{x}[c][c][1]{{\large $x$}} 
\psfrag{c}[c][r][1]{{\large $c$}}
\includegraphics[width=0.6\textwidth,height=7cm]{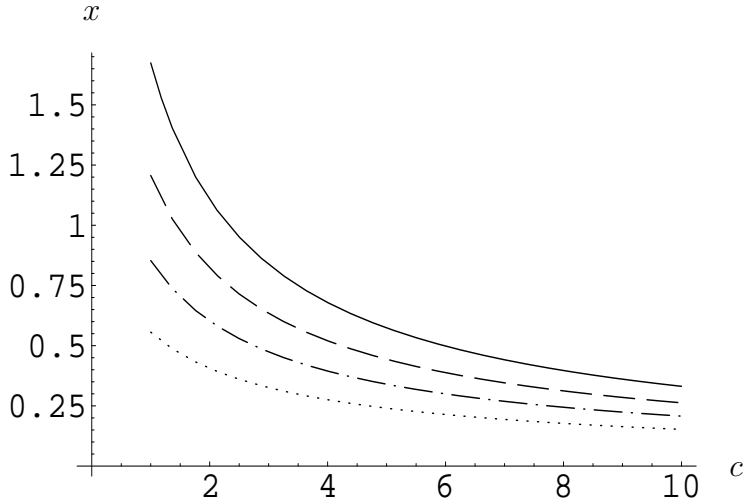}
\caption{Value of the upper bound on $x$ obtained in Eq.(\ref{FINAL}) as a function of 
parameter $c$, for $w\sqrt{kM_5}$ equal to $2 \mbox{TeV}$ (dotted line), $3 \mbox{TeV}$ 
(dot-dashed line), $5 \mbox{TeV}$ (dashed line) and $10 \mbox{TeV}$ (plain line).
The choice of those values for the parameter combination $w\sqrt{kM_5}$ is motivated 
by Eq.(\ref{wkM5}) and Eq.(\ref{wkM5bis}). The regions situated above the curves are
rejected by bound (\ref{FINAL}).} 
\label{fig:x}
\end{center}
\end{figure}

$\bullet$ {\bf Numerical values:}
Let us present and discuss the values for limit (\ref{FINAL}) on the $x$ parameter
of the $''+-+''$ model. We recall that this limit is nothing else but an expression 
of the experimental bound (\ref{thetaBound}) originating from considerations on the 
number of neutrino families. 
\\ In Fig.(\ref{fig:x}), we show the value of this limit (\ref{FINAL}) on $x$ as
a function of the parameter $c$. The other quantity $w\sqrt{kM_5}$, on which also
depends the limit (\ref{FINAL}), has been set around the TeV scale in this figure.
This choice is motivated by Eq.(\ref{wkM5bis}) which results from a combination of 
various constraints.
\\ The behavior of curves drawn on Fig.(\ref{fig:x}) can be explained physically 
in the following terms. 
The decrease of $c$ has two dominant effects on matrix 
(\ref{MassMatrix}). The first one is a decrease of the masses (\ref{Mnu1})-(\ref{MnuN}) 
mixing the SM neutrino $\nu_L$ with KK excitations of bulk neutrino $\psi^{(n)}$ 
[$n \geq 1$]. The second one is that the KK mass (\ref{mKK1}) for first excited mode 
$\psi^{(1)}$ increases, so that this excited state $\psi^{(1)}$ tends to decouple.
These two effects induce a decrease of the mixing, quantified typically by $\tan^2\theta_\nu$
(see Section \ref{Measure}), between the SM neutrino $\nu_L$ and sterile neutrinos 
$\psi^{(n)}$ [$n \geq 1$]. 
Therefore, a $c$ decrease can be compensated by an increase of $x$ in a way such that 
$\tan^2\theta_\nu$ remains fixed at a given value, since $\tan^2\theta_\nu$ increases 
with $x$ (see Eq.(\ref{thetaB})) as we have explained at the end of Section 
\ref{implication}.
This feature allows to understand why in Fig.(\ref{fig:x}) the $x$ value, which 
is associated to a value of $\tan^2\theta_\nu$ fixed to its limit: $0.0077$ (as the 
$x$ bound represented in Fig.(\ref{fig:x}) expresses the bound (\ref{thetaBound}) on 
$\tan^2\theta_\nu$), increases when $c$ diminishes, all other fundamental parameters 
($w$, $k$ and $M_5$) being fixed.

\vskip .5cm

\noindent $\bullet$ {\bf Bound on $x$ from the combination of constraints 
on $c$, $k$ and $x$:}
Motivated by the constraint on $k$ obtained in Eq.(\ref{Kbound}) and the condition
(\ref{GaugeHier}) concerning gauge hierarchy, we choose to consider Eq.(\ref{climB})
as the relevant constraint on $c$ originating from experimental bounds on neutrino 
masses. As it is clear from Fig.(\ref{fig:x}), by combining this constraint (\ref{climB}) 
on $c$ with the constraint (\ref{FINAL}) on $x$, we can obtain a value for the limit
on $x$ as a function of the quantity $w\sqrt{kM_5}$: the associated numerical results 
are respectively (the necessary condition $w\sqrt{kM_5} = {\cal O}(\mbox{TeV})$ is due 
to the relation obtained in Eq.(\ref{wkM5bis})), 
\begin{equation}
x \lesssim \{0.51;0.78;1.09;1.50\}
\ \ \ \mbox{for} \ \ 
w\sqrt{kM_5}=\{2;3;5;10\} 
\ \mbox{TeV}.
\label{FINALnum}
\end{equation}
Let us comment about the use of constraint (\ref{climB}) on $c$ for deriving
the limit (\ref{FINALnum}) on $x$. The constraint (\ref{climB}) on $c$
corresponds to the following exact values for the relevant fundamental parameters:
$w=10^{-15}$ and $k/M_5=1$. Now, only the associated orders of magnitude are 
imposed for $w$ and $k/M_5$ by the condition (\ref{GaugeHier}) concerning
gauge hierarchy and the constraint on $k$ obtained in Eq.(\ref{Kbound}), 
respectively. Hence, one may think of considering the constraint (\ref{climB})
on $c$ induced by other values of $w$ and $k/M_5$ around respectively $10^{-15}$ and $1$. 
However, the constraint (\ref{climB}) on $c$ possesses a weak dependence on both the
parameter $w$ (see end of Section \ref{PARAMc}) and ratio $k/M_5$ (see also 
Eq.(\ref{climA})). In conclusion, the dependences of $c$ constraint
(\ref{climB}) on $w$ and $k/M_5$ do not introduce another significant dependence 
on fundamental parameters for the $x$ limit (\ref{FINALnum}) (compared to the
dependence of $x$ limit (\ref{FINALnum}) on $w\sqrt{kM_5}$). 
\\ In summary, we have derived the experimental bound (\ref{FINALnum}) on
the fundamental parameter $x$ of the $''+-+''$ scenario. The obtained values 
of this upper bound (\ref{FINALnum}) are valid for $w \sim 10^{-15}$ 
(necessary for solving the gauge hierarchy problem), $k < M_5$ (condition 
(\ref{trust}) of validity for the $''+-+''$ model) with $k \sim M_5$ 
(resulting from various relevant constraints) and $M_5^3 \simeq k M_{Pl}^2$
(good approximation of relation (\ref{+-+krelat}) characteristic of the
$''+-+''$ framework), the two latter conditions leading to: 
$k \sim M_5 \sim M_{Pl}$.

From a general point of view, it is particularly interesting to obtain an 
experimental limit (as in Eq.(\ref{FINALnum})) on the fundamental parameter 
$x$ in the $''+-+''$ framework (with an additional massive bulk neutrino). 
As a matter of fact, among the five fundamental parameters of the $''+-+''$ 
model including a massive bulk neutrino, namely $x$, $w$, $k$, $M_5$ and $c$,
only $x$ is really free from the theoretical point of view. In the sense that
the four other fundamental parameters undergo the following direct constraints. 
The warp factor $w$ has to be approximately equal to $10^{-15}$ if the gauge
hierarchy question is to be addressed (see Eq.(\ref{GaugeHier})). The $AdS_5$ 
curvature $k$ must be of the same order of magnitude as the fundamental Planck 
mass $M_5$ in order to avoid the appearance of a hierarchy between energy scales. 
The value of gravity scale $M_5$ is restricted via the formula (\ref{+-+krelat}) 
(given in a good approximation by Eq.(\ref{M5exp})), which is dictated by the 
$''+-+''$ theory, to be a known function of the other fundamental parameters 
$x$, $w$ and $k$ (or equivalently $L_-$, $L_+$ and $k$). Finally, the quantity 
$c$, which parameterizes the amplitude of 5-dimensional neutrino mass $m$ (see
Eq.(\ref{VEV})), fixes the neutrino mass scales, and can thus be constrained by 
considering their realistic value (see Section \ref{PARAMc}).

\vskip .5cm

\noindent $\bullet$ {\bf Consequences of $x$ constraint for other parameters/quantities:}
The parameters $x$ (defined by Eq.(\ref{xDEF})) and $w$ (defined by Eq.(\ref{WarpFactor})) of the
$''+-+''$ scenario can be replaced by the theoretically equivalent parameters $L_-$ and $L_+$ 
(defined in Section \ref{Formalism}). Therefore, the experimental bound on $x$ obtained in 
Eq.(\ref{FINALnum}) together with the condition (\ref{GaugeHier}) ($\Leftrightarrow$ 
(\ref{GaugeHierBis})) on $w$ concerning the gauge hierarchy give rise to the approximative 
expression for $L_-$ and $L_+$:  
\begin{equation}
L_- \sim L_+ \sim 34/k. 
\label{L1L2}
\end{equation}

\begin{figure}[t]
\begin{center} 
\psfrag{m1}[c][c][1]{{\large $m^{(1)}_{KK}$ (TeV)}} 
\psfrag{c}[c][r][1]{{\large $c$}}
\includegraphics[width=0.6\textwidth,height=6cm]{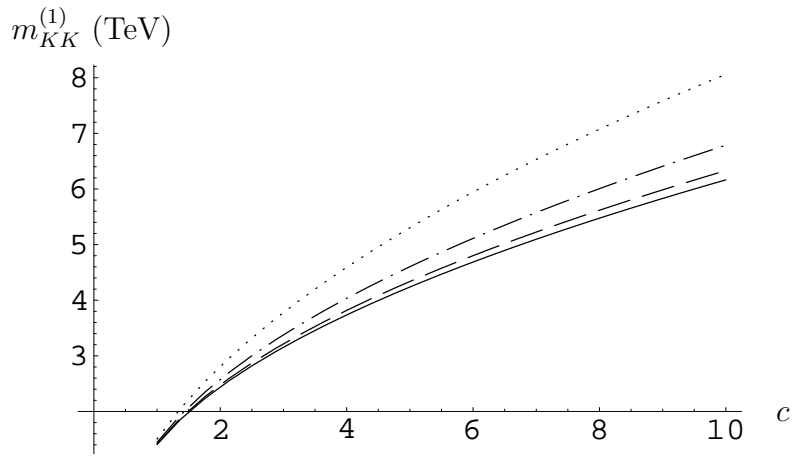}
\caption{Mass (\ref{mKK1}) for first neutrino KK excitation as a function of the parameter $c$.
The parameter $x$ in mass expression (\ref{mKK1}) has been set to its limit (\ref{FINAL}). The
quantity $w\sqrt{kM_5}$ entering Eq.(\ref{FINAL}) has been fixed to $2 \mbox{TeV}$ (dotted line), 
$3 \mbox{TeV}$ (dot-dashed line), $5 \mbox{TeV}$ (dashed line) and $10 \mbox{TeV}$ (plain line).
The choice of taking systematically $w\sqrt{kM_5}={\cal O}(\mbox{TeV})$ is motivated by the relevant 
condition obtained in Eq.(\ref{wkM5bis}). Furthermore, we have considered the case where $k=M_5$
(recall that having $k \sim M_5$ avoids the presence of a new typical energy scale value) so 
that the quantity $w\sqrt{kM_5}$ is equal to the product $wk$ involved in KK mass expression 
(\ref{mKK1}). Domains lying below the curves are ruled out (see Eq.(\ref{mKK1}) and 
Eq.(\ref{FINAL})).}
\label{fig:m1}
\end{center}
\end{figure}

From another point of view, the experimental constraint (\ref{FINAL}) on the parameter $x$
(illustrated in Fig.(\ref{fig:x})), which results from a study on the number of neutrino 
generations, induces in particular a lower bound on the KK mass $m_{KK}^{(1)}$ for first excited 
mode of bulk neutrino since this mass depends on $x$ through formula (\ref{mKK1}). 
Following the presentation of Fig.(\ref{fig:x}), this lower bound is shown on Fig.(\ref{fig:m1}) 
in function of the parameter $c$ as the limit (\ref{FINAL}) on $x$ depends on $c$.
We see on Fig.(\ref{fig:m1}) that this lower bound on $m_{KK}^{(1)}$, once combined with 
the bound (\ref{climB}) on $c$ (due to experimental constraints on neutrino masses), gives 
rise to the conservative bound:
\begin{equation}
m_{KK}^{(1)} \gtrsim 1.6 \ \mbox{TeV}.
\label{KK1bound}
\end{equation}

\vskip .5cm

\begin{figure}[t]
\begin{center} 
\psfrag{x}[c][c][1]{{\large $x$}} 
\psfrag{c}[c][r][1]{{\large $c$}}
\includegraphics[width=0.6\textwidth,height=6cm]{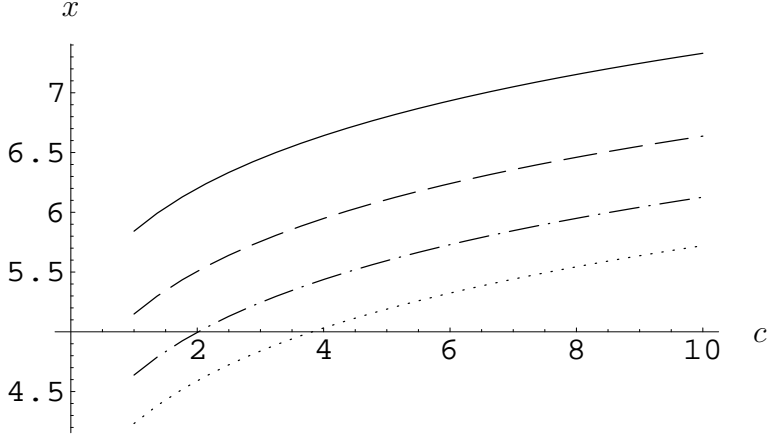}
\caption{Values of parameter $x$ (domains below the curves) corresponding to the situation where
only one neutrino mass eigenvalue is smaller than the $Z^0$ boson mass. Those values are presented
as a function of parameter $c$, for $wk$ equal to $2 \mbox{TeV}$ (dotted line), $3 \mbox{TeV}$ 
(dot-dashed line), $5 \mbox{TeV}$ (dashed line) and $10 \mbox{TeV}$ (plain line). We have also 
set the ratio $k/M_5$ at the physically relevant value of one so that each of the present curves 
(obtained for a fixed value of $wk$) can be associated to a curve in Fig.(\ref{fig:x})
(obtained for a fixed value of $w\sqrt{kM_5}$). This is useful as the domains shown here 
(below the curves) represent the regions of validity for the upper bounds on $x$ given in 
Fig.(\ref{fig:x}).}
\label{fig:domain}
\end{center}
\end{figure}

\noindent $\bullet$ {\bf Domain of validity for the obtained constraint on $x$:}
Let us consider the constraint (\ref{FINAL}) on $x$ which is illustrated in Fig.(\ref{fig:x}).  
This constraint was obtained from considerations on the $Z^0$ width measurements in the case
where only the lightest neutrino eigenstate $\nu_1$ has a mass smaller than the $Z^0$ mass 
(see Section \ref{Measure}). Therefore, this constraint holds in the region of parameter 
space where only one neutrino eigenstate is lighter than the $Z^0$ boson. This region is 
shown in Fig.(\ref{fig:domain}). It corresponds to the region in which all neutrino mass 
eigenvalues except the smallest one, namely $m_{\nu_2},m_{\nu_3},\dots$ (see foot-note \ref{index}), 
are larger than the $Z^0$ mass. We mention that those eigenvalues are approximatively given
by $m_{\nu_2} \simeq m_\nu^{(1)}$ and $m_{\nu_i} \simeq m_{KK}^{(i-1)}$ [$i \geq 3$] (see 
mass definitions in Section \ref{NeutMassMat}) for $c \in [1,10]$, $wk \in [2,10] \mbox{TeV}$ 
and $x \gtrsim 4$.
\\ We deduce from Fig.(\ref{fig:x}) and Fig.(\ref{fig:domain}) that we have only excluded 
intermediate values of $x$, or more precisely that the obtained range of values is,
\begin{equation}
x < {\cal O}(1)
\ \mbox{or} \ 
{\cal O}(4) \leftrightarrow {\cal O}(6) < x.
\label{B}
\end{equation}

We discuss now the scenario where there exist several neutrino mass eigenvalues
smaller than the $Z^0$ mass, a scenario which arises within the regions of parameter
space lying above the curves shown in Fig.(\ref{fig:domain}). For instance, let us consider
the simplest case with two neutrino mass eigenvalues smaller than the $Z^0$ mass. Then,
the admixture $\cos^2\theta_\nu$ introduced in Section \ref{Measure} would be now defined as,
\begin{equation}
\cos^2\theta_\nu=U^2_{01}+U^2_{02}, 
\label{DEFtwo}
\end{equation}
instead of Eq.(\ref{DEF}). In order to deduce a constraint on the parameters of $''+-+''$ model
from the bound (\ref{thetaBound}) on the angle $\theta_\nu$ due to $Z^0$ width measurements, one 
has to express $\theta_\nu$ (defined now by Eq.(\ref{DEFtwo})) in terms of those parameters. 
This is far from being trivial for the two following reasons. First, the squared matrix element 
$U^2_{02}$ entering Eq.(\ref{DEFtwo}) involves an unknown sum (see Eq.(\ref{U0I})) of 
Bessel function roots (via the KK masses $m_{KK}^{(m)}$ [$m \geq 1$] given in 
Eq.(\ref{mKK1})-(\ref{mKKN})). Secondly, the squared matrix element $U^2_{02}$ involves
the eigenvalue $m_{\nu_2}$ (see Eq.(\ref{U0I})), of the infinite neutrino mass matrix 
(\ref{MassMatrix}), which has to be expressed in terms of the mass matrix elements. 
Hence, in the domains of parameter space situated above the curves of Fig.(\ref{fig:domain}),
it turns out to be difficult to derive analytically a constraint on the parameter $x$ 
from the bound (\ref{thetaBound}) due to $Z^0$ width measurements. A numerical approach 
does not seem easier since the definition of admixture $\cos^2\theta_\nu$ depends on the 
number of neutrino mass eigenvalues smaller than the $Z^0$ mass (as exhibits well 
Eq.(\ref{DEFtwo})) so that the computation of constraint on $x$ depends strongly on the 
region of parameter space studied. Nevertheless, one can predict that this constraint on $x$,
issued from the bound (\ref{thetaBound}) on $\theta_\nu$, is less severe than the constraint 
presented in Fig.(\ref{fig:x}) (obtained in the domains below curves of Fig.(\ref{fig:domain})) 
since there are more positive contributions to the admixture $\cos^2\theta_\nu$ (compare 
Eq.(\ref{DEF}) with Eq.(\ref{DEFtwo}) for example).

\vskip .5cm

\begin{table}[!ht]
\begin{center}
\begin{tabular}{c|c|c}
& $w \sqrt{kM_5}=2 \ \mbox{TeV}$ & $w \sqrt{kM_5}=5 \ \mbox{TeV}$  \\ 
\hline
& {\bf [}${\bf k=3.82 \ 10^{17}}${\bf GeV]} & {\bf [}${\bf k=1.51 \ 10^{18}}${\bf GeV]}  \\
& & \\
$w=4.5 \ e^{-35}$ & $x \in [0,0.3]$ or $x \gtrsim 4.9$ & $x \in [0,0.6]$ or $x \gtrsim 6.3$ \\
$\ \ \ \ \ \ =2.84 \ 10^{-15}$ & & \\
& $x \in [0,1.2]\mbox{U}[2.3,3.1]\mbox{U}[4.2,16.6]$ & $x \in [0,1.8]\mbox{U}[3.1,17.4]$ \\
& & \\
\hline
& {\bf [}${\bf k=1.15 \ 10^{17}}${\bf GeV]} & {\bf [}${\bf k=4.56 \ 10^{17}}${\bf GeV]} \\
& & \\
$w=10 \ e^{-35}$ & $x \in [0,0.3]$ or $x \gtrsim 4.5$ & $x \in [0,0.6]$ or $x \gtrsim 5.9$ \\
$\ \ \ \ \ \ \ =6.30 \ 10^{-15}$ & & \\
& $x \in [0,1.1]\mbox{U}[2.2,2.6]\mbox{U}[3.2,16.5]$ & $x \in [0,1.8]\mbox{U}[2.8,17.2]$  
\end{tabular}
\caption{Experimental constraints on the parameter $x$ of the $''+-+''$ model issued from considerations
on neutrino physics [upper line] ({\it c.f.} Fig.(\ref{fig:x}) and Fig.(\ref{fig:domain})) 
and exchange of graviton KK modes [lower line] \cite{+-+A,+-+B} for $c=3$ and 
various values of the other theoretical parameters $w$ and $k$ ($M_5$ 
being fixed by the characteristic relation (\ref{+-+krelat}) or equivalently (\ref{+-+krelatWX})). The 
choice of restricting the parameter space to $w \sim 10^{-15}$ and $k \sim M_{Pl}$ is motivated, 
respectively, by the gauge hierarchy question (see Section \ref{Formalism}) and the condition $k \sim M_5$ 
(leading together with Eq.(\ref{+-+krelatWX}), or equivalently Eq.(\ref{M5exp}), to $k \sim M_{Pl}$) 
under which no new typical energy scale value is introduced. Finally, all parameter values are taken 
such that the necessary condition $k < M_5$ (see Eq.(\ref{trust})) is well fulfilled.}
\label{MyTable}
\end{center}
\end{table}

\noindent $\bullet$ {\bf Comparison of the bound placed on $x$ with existing bounds:}
We now compare our experimental bound on $x$ obtained in Fig.(\ref{fig:x}) 
(holding in the parameter spaces illustrated on Fig.(\ref{fig:domain})) with 
the other experimental bounds on $x$ which have already been derived in the literature
\cite{+-+A,+-+B} within the $''+-+''$ framework. These other bounds on $x$ have been
obtained by considering KK excitations of the graviton. More precisely, the authors of 
\cite{+-+A,+-+B} have placed a constraint on $x$ by requiring that the contribution of
resonant graviton KK state production to the SM process $e^+e^-\to\mu^+\mu^-$ is not
visible at leptonic colliders. Furthermore, another bound on $x$ has been deduced from
the condition that the exchange of graviton KK modes does not induce noticeable 
corrections to the Newton law (which is tested experimentally) \cite{+-+A,+-+B}.
\\ First, we consider our upper bound (valid for $x \lesssim 4.9$):
\begin{equation}
x \lesssim 0.3,
\label{Validity}
\end{equation}
obtained (see Fig.(\ref{fig:x})) for the typical value $c=3$ and,
\begin{equation}
w \sqrt{kM_5}=2 \ \mbox{TeV}.
\label{Valid}
\end{equation}
Let us also consider, for example, the precise warp factor value (bound (\ref{Validity}) 
is valid for $w \sim 10^{-15}$):
\begin{equation}
w=4.5 \ e^{-35}=2.84 \ 10^{-15}.
\label{Val}
\end{equation}
For values of $x$ and $w$ given respectively by Eq.(\ref{Validity}) and 
Eq.(\ref{Val}), the property (\ref{+-+krelatWX}) of $''+-+''$ scenario reads 
in a good approximation as $M_5^3 \simeq k M_{Pl}^2$, a relation which leads together 
with Eq.(\ref{Valid}) and Eq.(\ref{Val}) to, 
\begin{equation}
k=3.82 \ 10^{17} \ \mbox{GeV}.
\label{Va}
\end{equation}
To summarize, our bound (\ref{Validity}) on $x$ holds for the values of fundamental 
parameters $w$, $k$ and $M_5$ given respectively by Eq.(\ref{Val}), 
Eq.(\ref{Va}) and Eq.(\ref{+-+krelat}) ($\Leftrightarrow$ Eq.(\ref{+-+krelatWX})).
For the same values of parameters $w$, $k$ and $M_5$, the bound on $x$, coming from 
considerations on graviton KK state production at colliders, is \cite{+-+A,+-+B},  
\begin{equation}
x \in [0,1.2]\mbox{U}[2.3,3.1], \ \mbox{or}, \ x \gtrsim 4.2,
\label{Colliders}
\end{equation}
while the bound due to possible modifications of gravity reads as \cite{+-+A,+-+B},
\begin{equation}
x \lesssim 16.6
\label{Gravity}
\end{equation}
All those bounds on $x$ are summarized in Table \ref{MyTable} together with other values
of the bounds associated to different parameter values
\footnote{For the values of warp factor $w$ considered in Table \ref{MyTable}, 
no additional bound on $x$ can be put by considering the production of graviton 
KK mode at leptonic colliders (via the reaction $e^+e^-\to\gamma + light \ KK \ mode$
giving a missing energy signal) \cite{+-+A,+-+B}.}. 
The results presented in Eq.(\ref{Validity}), Eq.(\ref{Colliders}) and 
Table \ref{MyTable} show that our constraint on $x$, obtained from the study of
experimental bounds on neutrino masses, is typically stronger than the existing  
constraint coming from collider physics.

\subsubsection{The case of three lepton flavors}
\label{3Flavors}

\noindent $\bullet$ {\bf Generalization of the neutrino mass terms:}
Here, we discuss the realistic situation where there are three families of neutrino. 
First, in this case, the neutrino mass matrix (\ref{MassMatrix}) must be modified.
As a matter of fact, 
it is natural to assign 3 different 5-dimensional masses $m_f$ [$f=e,\mu,\tau$] 
(see Eq.(\ref{Action})), and thus 3 parameters $c_f$ (see Eq.(\ref{VEV})), to the 3 
generations of bulk neutrino $\Psi_f$. Therefore, there are 3 masses $m_{\nu_f}^{(n)}$ 
[$f=e,\mu,\tau;n=0,1,2,\dots$], mixing the right-handed modes of 3 bulk neutrinos 
$\psi^{(n)}_{fR}$ with the 3 left-handed SM neutrinos $\nu_{fL}$, which are associated 
to the 3 parameters $c_f$ (see Eq.(\ref{Mnu0})-(\ref{MnuN})). While this mass 
$m_{\nu_f}^{(n)}$, which must enter Eq.(\ref{MassMatrix}), differs for each family of 
bulk neutrino state $\psi^{(n)}_{fR}$ (associated to $c_f$), it is identical for each 
family of SM neutrino $\nu_{fL}$ if one assumes a universal value for the Yukawa coupling 
parameter $\lambda_5$ (see the action (\ref{Action})). Similarly, there are now 3 types 
of KK mass $m_{KKf}^{(n)}$ [$f=e,\mu,\tau;n=1,2,3,\dots$], for the excited modes 
of 3 bulk neutrinos $\psi_f^{(n)}$, which correspond to the 3 parameters $c_f$ 
(see Eq.(\ref{mKK1})-(\ref{mKKN})).

\vskip .5cm

\noindent $\bullet$ {\bf New bound on $x$:}
In Section \ref{PARAMx}, we deduced the bound (\ref{FINALnum}) on parameter $x$ from 
the constraints (\ref{FINAL}) (due to measurements of the $Z^0$ width: see Section \ref{Ngen}) 
and (\ref{FINALbis}) (due to data on neutrino masses: see Section \ref{NeutMass}) in the simplified 
case of 1 lepton family. 
Let us discuss the changes of those limits (\ref{FINALnum}), (\ref{FINAL}) and (\ref{FINALbis}) 
as one passes to the case of 3 lepton families.

First, we determine the equivalent of constraint (\ref{FINAL}) in the case of 3 lepton species.
By extending the calculation performed in Appendix \ref{ApMix} to $N_f=3$ flavors, we obtain 
the following expression for the neutrino mixing angle $\theta_\nu$, which enters the 
experimental constraint (\ref{thetaBound}),
\begin{equation}
\tan^2\theta_\nu \simeq
\frac{1}{N_f}-1+ 
\sum_{m=1}^\infty \bigg ( \frac{m_{\nu_e}^{(m)}}{m_{KKe}^{(m)}} \bigg )^2 +
\sum_{n=1}^\infty \bigg ( \frac{m_{\nu_\mu}^{(n)}}{m_{KK\mu}^{(n)}} \bigg )^2 +
\sum_{p=1}^\infty \bigg ( \frac{m_{\nu_\tau}^{(p)}}{m_{KK\tau}^{(p)}} \bigg )^2.
\label{wanted3F}
\end{equation}
We notice that, for $N_f=1$ flavor, this formula reduces well to the result (\ref{wanted})
found in Appendix \ref{ApMix}. After replacing the masses $m_{\nu_f}^{(n)}$ and $m_{KKf}^{(n)}$ 
[$f=e,\mu,\tau;n=1,2,3,\dots$], entering Eq.(\ref{wanted3F}), by their own expression (see the
above discussion, Eq.(\ref{Mnu1})-(\ref{MnuN}) and Eq.(\ref{mKK1})-(\ref{mKKN})), we find (in
the same way as in Section \ref{implication}):
\begin{equation}
\tan^2\theta_\nu \simeq 
\frac{1}{N_f}-1+ 
\frac{\upsilon^2}{w^2kM_5}
\sum_{f=e,\mu,\tau}
\bigg [ \frac{e^{(2c_f+1)x}}{2(2c_f+1)}
+ g(c_f) \bigg ],
\label{thetaB3F} 
\end{equation}
the function $g$ being defined as in Eq.(\ref{thetaB}). The $\tan^2\theta_\nu$ expression 
(\ref{thetaB3F}) for 3 flavors exhibits the same kind of structure and same dependence on fundamental 
parameters ($x$, $w$, $k$, $M_5$ and $c_f$ [$f=e,\mu,\tau$]) as the $\tan^2\theta_\nu$ expression 
(\ref{thetaB}) for 1 flavor. Hence, for identical parameter values, one expects the limit (\ref{FINAL}) 
for 1 flavor (value given in Fig.(\ref{fig:x})), 
derived from the combination of constraint (\ref{thetaBound}) on $\tan^2\theta_\nu$ (due to $Z^0$ width 
measurements) with the $\tan^2\theta_\nu$ expression, to be of the same order of magnitude as the
equivalent limit for 3 flavors.

Secondly, we comment the constraint (\ref{FINALbis}) in the situation where 3 lepton species 
are considered. In such a situation, the experimental limits on absolute physical neutrino masses
are of order of the eV like in the case of a unique lepton generation (first one), as shown in Section 
\ref{NeutMass}. By consequence, one expects that, for 3 generations, imposing these experimental limits 
(on the 3 smallest neutrino mass eigenvalues $m_{\nu_{1,2,3}}$) leads to bounds on the 3 parameters 
$c_f$ [$f=e,\mu,\tau$] of same order as the bound on $c$ (given in Eq.(\ref{climA})-(\ref{climB})) 
obtained for 1 generation from the constraint (\ref{FINALbis}) (application of the experimental limit 
on the smallest neutrino mass eigenvalue $m_{\nu_1}$).

Finally, since the values of limit (\ref{FINAL}) (due to $Z^0$ width measurements) and limit on $c$ 
from constraint (\ref{FINALbis}) (due to data on neutrino masses) are expected to remain of the same 
order of magnitude when one passes to the case of 3 lepton flavors (see above discussions), the bound 
(\ref{FINALnum}) on parameter $x$, which is deduced from those two limits, should also still be of the 
same order in the case of 3 lepton flavors.

Similarly, in the case of 3 lepton flavors, by determining the parameter space where 3 neutrino mass 
eigenvalues are smaller than the $Z^0$ mass, one should find a domain of validity for the limit due to
$Z^0$ width measurements that resemble the domain shown in Fig.(\ref{fig:domain}) for the 1 flavor 
case. Indeed, the neutrino mass matrix elements have the same dependence on the fundamental parameters 
in both the cases of 1 and 3 flavors.

\vskip .5cm

\noindent $\bullet$ {\bf Characteristic examples:}
Now, we will give examples of values, within the case of 3 lepton flavors, for the bound on parameter 
$x$ deduced from the constraints due to $Z^0$ width measurements and experimental limits on absolute 
neutrino masses. Those typical examples will confirm the expectation (see above) that this bound on 
$x$ has the same order of magnitude in the two cases of 1 and 3 flavors.

Let us start by considering the simplified scenario where the 3 parameters $c_f$ 
[$f=e,\mu,\tau$] are related, for instance, through the formula,
\begin{equation}
c_\tau=1.5 \ c_\mu=(1.5)^2 \ c_e,
\label{ex3FIa} 
\end{equation}
which reduces the number of degrees of freedom. The hypothesis (\ref{ex3FIa}) is motivated by the 
fact that values, for the 3 fundamental parameters $c_f$, of similar orders of magnitude are desirable.
Under this assumption (\ref{ex3FIa}), requiring that the 3 smallest eigenvalues $m_{\nu_{1,2,3}}$ of 
neutrino mass matrix (\ref{MassMatrix}) (modified to involve the 3 $c_f$ parameters) are smaller than 
the eV scale (see Section \ref{NeutMass}) yields the numerical result: 
\begin{equation}
c_e \gtrsim 1.25 \ \ \ \mbox{for} \ k/M_5=1 \ \mbox{and} \ w=2\mbox{TeV}/M_{Pl},
\label{ex3FIb} 
\end{equation}
\begin{equation}
c_e \gtrsim 1.29 \ \ \ \mbox{for} \ k/M_5=1 \ \mbox{and} \ w=10\mbox{TeV}/M_{Pl},
\label{ex3FIbBIS} 
\end{equation}
which is close to the same result obtained in the case of 1 neutrino
generation for similar values of $w$ (see Eq.(\ref{climB}) with following text). 
Then, by taking into account this bound (\ref{ex3FIb})-(\ref{ex3FIbBIS}) on $c_e$ (together with 
Eq.(\ref{ex3FIa})) and applying the constraint (\ref{thetaBound}) (from $Z^0$ width measurements) 
on the $\tan^2\theta_\nu$ expression (\ref{thetaB3F}) in terms of all the fundamental parameters, we 
find (with $k = M_5 = M_{Pl}$ accordingly to characteristic relation (\ref{M5exp})),
\begin{equation}
x \lesssim \{1.02;1.00\}
\ \ \ \mbox{for} \ \ 
w\sqrt{kM_5}=\{2;10\} 
\ \mbox{TeV}.
\label{ex3FIc}
\end{equation}

The other example of simplification hypothesis we consider, namely,
\begin{equation}
c_\tau=2.5 \ c_\mu=(2.5)^2 \ c_e,
\label{ex3FIIa} 
\end{equation}
corresponds to values of the 3 $c_f$ parameters more distinct than in the first hypothesis 
(\ref{ex3FIa}). Under the assumption (\ref{ex3FIIa}), the constraints from $Z^0$ width measurements 
(namely the constraint (\ref{thetaBound}) on $\tan^2\theta_\nu$) and from limits on neutrino mass scales 
(given in Section \ref{NeutMass}) lead to the same bound on $c_e$ as in
Eq.(\ref{ex3FIb})-(\ref{ex3FIbBIS}) and then, by using expression (\ref{thetaB3F}), to (with 
$k = M_5 = M_{Pl}$):
\begin{equation}
x \lesssim \{0.47;0.46\}
\ \ \ \mbox{for} \ \ 
w\sqrt{kM_5}=\{2;10\} 
\ \mbox{TeV}.
\label{ex3FIIb}
\end{equation}

In conclusion, the characteristic values (\ref{ex3FIc}) and (\ref{ex3FIIb}) for the bound 
on parameter $x$ (issued from experimental considerations on neutrino physics), which were 
derived in the case of 3 neutrino flavors, are of same order as the values for the identical 
bound obtained in the case of 1 neutrino flavor (see Eq.(\ref{FINALnum})).

\vskip .5cm

\noindent $\bullet$ {\bf Additional experimental constraints:}
In Section \ref{Main}, we have used some experimental data on neutrino physics
(coming from measurements of $Z^0$ width and effective neutrino masses) in order to
constrain the $''+-+''$ model, and in particular to place a bound on the
parameter $x$, within the typical case of 1 lepton flavor. 
\\ In the more precise 
case of 3 lepton flavors, one could also use the bounds on squared neutrino mass 
differences ($\Delta m^2_{32}$ and $\Delta m^2_{21}$) and lepton mixing angles
($\theta_{12}$, $\theta_{23}$ and $\theta_{13}$), which are derived from the 
present results of neutrino flavor oscillation experiments \cite{SquaDiff,lastKAMLAND2F}. 
Nevertheless, these lepton mixing angles depend on the neutrino mass matrix, 
which is predicted by the considered $''+-+''$ scenario (with additional 
massive bulk neutrinos), but also on the charged lepton mass matrix, which in 
contrast is not determined by our scenario. Therefore, in order to use the 
bounds on these lepton mixing angles for constraining the $''+-+''$ model, 
one should consider a complementary model dictating the flavor structure of charged 
lepton masses, which is beyond the scope of our study. 
\\ We also mention that, in a detailed approach based on 3 lepton species,
one could also envisage to use the astrophysical and cosmological constraints (like
those due to considerations on big bang nucleosynthesis and duration of the
supernova 1987A neutrino burst \cite{AstroCosmo}) on mixing angles between a sterile 
neutrino and an active SM neutrino (either $\nu_e$, $\nu_\mu$ or $\nu_\tau$), as 
well as the SNO (salt phase) data \cite{Paola,SNO} on fraction of sterile neutrino 
component in the resultant solar $\nu_e$ flux at Earth, and, the experimental bound on 
branching ratio of the flavor violating decay channel $\mu \to e \gamma$ (which is
enhanced by the presence of significantly massive sterile neutrinos \cite{Kitano}). 
As a matter of fact, remind that, within the considered $''+-+''$ scenario, the KK 
excitations of bulk neutrinos (denoted as $\psi_f^{(n)}$ [$f=e,\mu,\tau;n \geq 1$]) 
behave like sterile neutrinos.

\section{The cases of the $''++''$ and $''++-''$ models}
\label{++AND++-}

\subsection{The $''++''$ model}

Here, we discuss, within the same philosophy as above, the case of another realization 
of the paradigm of anomalously light KK excitations: the $''++''$ model.
\\ The absence of any $''-''$ brane in the $''++''$ model protects it against the problem 
of radion ghost fields. Nevertheless, it turns out that for the construction of such 
a $''++''$ configuration, it is essential to have $AdS_4$ geometry on both branes 
\cite{++A,++B}.

In the $''++''$ model, the warp function $e^{-\sigma(y)}$ (which determines the 
metric as shown in Eq.(\ref{RSmetric})
\footnote{In the $''++''$ context, with our definition (\ref{RSmetric}) of the metric, 
the warp function $e^{-\sigma(y)}$ must also depend on the familiar coordinates 
$x^\mu$ [$\mu=1,\dots,4$] \cite{++A,++B}.}) 
can reach a minimum at a point $y=y_0$ lying between 
the two $''+''$ branes which sit on the two orbifold fixed points at $y=0$ and $y=L_+$ 
(where we live) \cite{++A,++B}. This geometrical feature is also a fundamental 
characteristic of the $''+-+''$ model in which the warp function $e^{-\sigma(y)}$ 
reaches also a minimum at a point $y=L_-$ (see Eq.(\ref{WarpFunction})) lying between two 
$''+''$ branes at $y=0$ and $y=L_+$ (here, there exist a $''-''$ brane at the extremum 
$y=L_-$). In this sense, the geometrical configuration of the $''++''$ model mimics
that of the $''+-+''$ model.  
\\ This similarity has two main consequences. First, the localizations of bulk
neutrino KK modes, and thus the masses (given by Eq.(\ref{Mnu0})-(\ref{MnuN})
in the $''+-+''$ model) mixing those KK 
modes with SM neutrinos (trapped at $y=L_+$), should be comparable in the $''++''$ 
and $''+-+''$ models. Secondly, the KK masses (given in
Eq.(\ref{mKK1})-(\ref{mKKN}) for $''+-+''$ and in \cite{Mouslopoulos} for $''++''$) 
for excitations of bulk neutrinos are similar (with identical dependences on
theoretical parameters) in the $''++''$ and $''+-+''$ models.
\\ Therefore, the whole effective neutrino mass matrix (which involves only these masses 
(\ref{Mnu0})-(\ref{MnuN}) and (\ref{mKK1})-(\ref{mKKN}), as shows Eq.(\ref{MassMatrix}), 
within the $''+-+''$ model) is expected to behave similarly in the equivalent parameter 
spaces of the $''++''$ and $''+-+''$ scenarios. Hence, in the $''++''$ scenario (for 
$w=w(k,L_+,y_0) \sim 10^{-15}$), one expects to deduce, from experimental constraints 
on neutrinos, limits on the equivalent parameter 
\begin{equation}
x=k(L_+-y_0)
\label{secondX}
\end{equation} 
of the same order as the limits on $x$ that we have obtained within the 
$''+-+''$ framework, namely:
\begin{equation}
x < {\cal O}(1)
\ \mbox{or} \ 
{\cal O}(4) \leftrightarrow {\cal O}(6) < x.
\label{secondB}
\end{equation}

\subsection{The $''++-''$ model}

Let us finally study the third multi-brane RS extension, namely the consistent
$''++-''$ model. Recall that the $''++-''$ model (see Section \ref{intro})
consists of a $''+''$ and a $''-''$ branes placed at the two orbifold fixed 
points $y=0$ and $y=\pi R_c$ respectively, the SM fields being confined on a 
second $''+''$ brane (at $y=L_+$) which moves freely in between ($0<L_+<\pi R_c$).
Hence, this model does not contain any freely moving $''-''$ brane and thus does not
give rise to the existence of radion ghost fields \cite{systradion}.

We begin by describing the neutrino mass matrix in the $''++-''$ framework. The 
elements $m_\nu^{(m)}$ (mixing the SM neutrinos with bulk neutrino KK excitations) 
of neutrino mass matrix (see Eq.(\ref{MassMatrix})) are given by \cite{Mouslopoulos},
\begin{equation}
m_\nu^{(0)} \simeq \sqrt{\frac{k_1}{M_5}(c-\frac{1}{2})} \ w^{c-1/2} \ \upsilon,
\label{Mnu0bis}
\end{equation}
\begin{equation}
m_\nu^{(n)} \simeq \frac{8 \ \zeta^{-\ 2}_n}{J_{c+1/2}(\zeta^-_n)} 
\ \bigg ( \frac{k_2}{k_1} \bigg ) ^{3/2} \sqrt{c} \ e^{-3x} \ \upsilon 
~~~~~~~[n=1,2,3,\dots],
\label{MnuNbis}
\end{equation}
where $k_1$ and $k_2$ are the two curvatures of the bulk (satisfying $k_1 \sim k_2$
but with $k_1 < k_2$), $w=w(k_1,L_+) \sim 10^{-15}$, $\zeta^-_n$ is the $n$-th root 
of $J_{c-1/2}(X)=0$ and (to be compared with parameters (\ref{xDEF}) and 
(\ref{secondX})),
\begin{equation}
x=k_2(\pi R_c-L_+).
\label{thirdX}
\end{equation}
We note that the mass $m_\nu^{(0)}$ for the 0-mode of neutrino given in 
Eq.(\ref{Mnu0bis}) has the same expression as in the $''+-+''$ context (see 
Eq.(\ref{Mnu0})).
\\ The other relevant elements of neutrino mass matrix (see Eq.(\ref{MassMatrix})),
namely the KK masses $m_{KK}^{(m)}$ (for bulk neutrino excitations), read as
\cite{Mouslopoulos},
\begin{equation}
m_{KK}^{(m)} = \zeta^-_m \ w \ e^{-(x+k_2L_+)} \ k_2
~~~~~~~[m=1,2,3,\dots].
\label{MKKbis}
\end{equation}

We now treat the bound (\ref{thetaBound}) on the angle $\theta_\nu$ issued from 
$Z^0$ width measurements within the $''++-''$ framework. The admixture 
$\cos^2\theta_\nu$ introduced in Section \ref{Measure} is defined by (see 
Eq.(\ref{U0I}) for the expression of $U^2_{0i}$),
\begin{equation}
\cos^2\theta_\nu=\sum^N_{i=1} U^2_{0i}, 
\label{DEFthree}
\end{equation}
where the index $i=1,2,\dots,N$ labels the $N$ neutrino eigenstates 
$\nu_1,\nu_2,\dots,\nu_N$ which are lighter than the $Z^0$ boson (see 
foot-note \ref{index}).
For $i \in [1,2,\dots,N]$, there exist an index $m_{min}$ such that for
$m>m_{min}$ the relation
\begin{equation}
\frac{m_\nu^{(m)\ 2}}{m_{KK}^{(m)\ 2}-m_{\nu_i}^2}
\ \frac{m_{KK}^{(m)\ 2}}{m_{KK}^{(m)\ 2}-m_{\nu_i}^2}
>
\frac{m_\nu^{(m)\ 2}}{m_{KK}^{(m)\ 2}}
\label{argument}
\end{equation}
becomes true (because the ratio $m_{KK}^{(m)\ 2}/m_{\nu_i}^2$ becomes 
sufficiently large). Now, the sum $\sum^\infty_{m=1}$ 
$m_\nu^{(m)\ 2}/m_{KK}^{(m)\ 2}$ 
is divergent. Indeed, this sum reads as (see Eq.(\ref{MnuNbis}) and Eq.(\ref{MKKbis})),
\begin{equation}
\sum^\infty_{m=1} \frac{m_\nu^{(m)\ 2}}{m_{KK}^{(m)\ 2}}
\simeq
\frac{64 \ c}{w^2} \ \frac{v^2\ k_2}{k_1^3} \ e^{2(k_2L_+-2x)} \ 
\sum^\infty_{m=1} \frac{\zeta^{-\ 2}_m}{J^2_{c+1/2}(\zeta^-_m)}, 
\label{dvgsum}
\end{equation}
and one has $\zeta^{-\ 2}_{m+1}>\zeta^{-\ 2}_m$ and 
$J^2_{c+1/2}(\zeta^-_{m+1})<J^2_{c+1/2}(\zeta^-_m)$.
Therefore, we deduce from Eq.(\ref{argument}) that all the sums entering ({\it c.f.} 
Eq.(\ref{U0I})) all the squared matrix elements $U^2_{0i}$ of Eq.(\ref{DEFthree})
diverge, so that all those elements tend to zero. By consequence, the 
phenomenological condition (\ref{thetaBound}) on $\theta_\nu$ ({\it c.f.} 
Eq.(\ref{DEFthree})) cannot be fulfilled within the $''++-''$ context.

\section{Conclusion}
\label{conclu}

We have studied the paradigm of anomalously light KK excitations through its different
realizations, namely the multi-brane RS extensions of type $''+-+''$, $''++''$ and $''++-''$.
The considered parameter space corresponds to the domains where the gauge hierarchy problem
is effectively solved. We have assumed that massive right-handed neutrinos (added to the SM)
propagate in the bulk.
\\ We have shown that the present experimental bounds on neutrino masses and mixing angles 
either constrain (see the respective limits (\ref{B}) and (\ref{secondB}) on geometrical 
parameters (\ref{xDEF}) and (\ref{secondX}) of the $''+-+''$ and $''++''$ models) relatively
strongly ({\it c.f.} Section \ref{PARAMx}) or exclude (as it occurs for the $''++-''$ model)
the theoretical realizations of the paradigm.

\vspace{1cm}

\noindent {\bf \Large Acknowledgments}

\noindent The author is grateful to J.-M.~Fr\`ere and A.~Papazoglou for 
stimulating conversations. It is also a pleasure to thank N.~Cosme, 
M.~Fairbairn, F.-S.~Ling, M.~Neubert and P.~Tiniakov for useful discussions.  
This work is supported by the Belgian Science Policy (IAP V/27) and the French 
Community of Belgium (IISN).

\newpage

\appendix
\noindent {\bf \Large Appendix}
\vspace{0.5cm}

\renewcommand{\thesubsection}{A.\arabic{subsection}}
\renewcommand{\theequation}{A.\arabic{equation}}
\setcounter{subsection}{0}
\setcounter{equation}{0}

\section{Neutrino mass eigenvalues}
\label{ApEigen}

Within the $''+-+''$ model,
the method we use in order to obtain the physical neutrino masses 
\footnote{The indexes $i$ of physical neutrino masses $m_{\nu_i}$ 
are chosen such that: $m_{\nu_1}<m_{\nu_2}<m_{\nu_3}\dots$
\label{index}} $m_{\nu_i}$ 
is to diagonalize the hermitian square of neutrino mass matrix 
${\cal M}$ (see Eq.(\ref{LagMass}) and Eq.(\ref{MassMatrix})): 
\begin{equation}
{\cal M}{\cal M}^\dagger=
U \ diag(m^2_{\nu_1},m^2_{\nu_2},m^2_{\nu_3},\dots) \ U^\dagger,
\label{DIAGO}
\end{equation}
$U$ being the neutrino mixing matrix defined by Eq.(\ref{UmatDEF}). We note that 
the Yukawa coupling constant $\lambda_5$ ({\it c.f.} Eq.(\ref{Action})) has been taken
real (see Section \ref{NeutMassMat}) so that the neutrino mass matrix ${\cal M}$ 
does not involve any CP violation phase. The presence of non-vanishing complex phases 
would not affect our study.
\vskip .4cm
$\bullet$ {\bf One KK excitation:}
\\ First, we observe that the mass ratio $m_\nu^{(0)}/m_\nu^{(1)}$, which is equal to 
$w^{c-1/2}$ (see Eq.(\ref{Mnu0}) and Eq.(\ref{Mnu1})), has a typical value much smaller 
than one. Indeed, the gauge hierarchy problem is solved for $w \sim 10^{-15}$ 
({\it c.f.} Eq.(\ref{GaugeHier})) and the experimental bounds on neutrino mass imply
$c \gtrsim 1.08$ (in a conservative way: see Section \ref{PARAMc}), which lead to 
$w^{c-1/2} \lesssim 2 \ 10^{-9}$.

In the simplified case where only the first KK excitation of bulk neutrino is considered, 
a straightforward calculation, at first order in $m_\nu^{(0)}/m_\nu^{(1)}$, gives us the 
following expressions for the two squared neutrino mass eigenvalues of 
${\cal M}{\cal M}^\dagger$,
\begin{equation}
m^2_{\nu_1} \simeq 
m_{KK}^{(1)\ 2} 
\ \frac{m_\nu^{(0)\ 2}}{m_{KK}^{(1)\ 2}+m_\nu^{(1)\ 2}},
\end{equation}
\begin{equation}
m^2_{\nu_2} \simeq m_{KK}^{(1)\ 2}+m_\nu^{(1)\ 2} \ 
\bigg (1+\frac{m_\nu^{(0)\ 2}}{m_{KK}^{(1)\ 2}+m_\nu^{(1)\ 2}}\bigg ).
\end{equation}    
\vskip .4cm  
$\bullet$ {\bf The KK tower:}
\\ Let us begin by determining the smallest squared neutrino mass eigenvalue
$m^2_{\nu_1}$, in the general case of an infinite tower of KK states. The eigenvalues
$m^2_{\nu_i}$, entering Eq.(\ref{DIAGO}), are solutions of the equation,
\begin{equation}
det [{\cal M}{\cal M}^\dagger-m^2_{\nu_i} {\bf 1}]=0,
\end{equation}
where ${\bf 1}$ denotes the identity matrix. After calculation of the determinant,
this equation can be rewritten as,
\begin{equation}
\bigg [\sum_{p=0}^\infty m_\nu^{(p)\ 2}-m^2_{\nu_i}-\sum_{p=1}^\infty 
\frac{m_\nu^{(p)\ 2}\ m_{KK}^{(p)\ 2}}{m_{KK}^{(p)\ 2}-m^2_{\nu_i}} \bigg ]
\prod_{p=1}^\infty (m_{KK}^{(p)\ 2}-m^2_{\nu_i})=0.
\label{ExplDet}
\end{equation}
Since taking $m^2_{\nu_i}=m_{KK}^{(p)\ 2}$ leads to a divergence 
in Eq.(\ref{ExplDet}), this equation is equivalent to,
\begin{equation}
\sum_{p=0}^\infty m_\nu^{(p)\ 2}-m^2_{\nu_i}-\sum_{p=1}^\infty 
\frac{m_\nu^{(p)\ 2}\ m_{KK}^{(p)\ 2}}{m_{KK}^{(p)\ 2}-m^2_{\nu_i}}=0,
\label{ExplDetBis}
\end{equation}
which can be transformed into,
\begin{equation}
m_\nu^{(0)\ 2}-m^2_{\nu_i}\bigg (1+\sum_{p=1}^\infty 
\frac{m_\nu^{(p)\ 2}}{m_{KK}^{(p)\ 2}-m^2_{\nu_i}}\bigg )=0.
\label{Trans}
\end{equation}
Assuming that an eigenvalue is much smaller than the weakest squared 
KK mass $m_{KK}^{(1)\ 2}$, one can deduce its expression at leading order 
from Eq.(\ref{Trans}) and the result is,
\begin{equation}
m^2_{\nu_1} \simeq \frac{m_\nu^{(0)\ 2}}{1+\sum_{p=1}^\infty 
\frac{m_\nu^{(p)\ 2}}{m_{KK}^{(p)\ 2}}}.
\label{ResultI}
\end{equation}
This eigenvalue would be the smallest one since the others are larger than
$m_{KK}^{(1)\ 2}$ (as we will see in Eq.(\ref{ResultII})).
In fact, the squared mass (\ref{ResultI}) is effectively the weakest eigenvalue
because the hypothesis made to derive it, namely,
\begin{equation}
m^2_{\nu_1} \ll m_{KK}^{(1)\ 2},
\label{APPROX}
\end{equation}
constitutes a good approximation in our framework (see the bounds on
$m_{\nu_1}$, in Eq.(\ref{TritiumApplied}), and on $m_{KK}^{(1)}$, in 
Eq.(\ref{KK1bound}) and Fig.(\ref{fig:m1})).

We discuss now the other squared neutrino mass eigenvalues $m^2_{\nu_i}$
[$i\geq 2$]. At leading order in $m_\nu^{(0)}/m_\nu^{(1)}$, 
Eq.(\ref{ExplDetBis}) becomes, 
\begin{equation}
m^2_{\nu_i} \bigg (1+\sum_{p=1}^\infty 
\frac{m_\nu^{(p)\ 2}}{m_{KK}^{(p)\ 2}-m^2_{\nu_i}}\bigg )=0.
\label{Others}
\end{equation}
The solution $m^2_{\nu_i}=0$ of this equation corresponds, at leading order in 
$m_\nu^{(0)}/m_\nu^{(1)}$, to the eigenvalue $m^2_{\nu_1}$ given by
Eq.(\ref{ResultI}). The other solutions $m^2_{\nu_i}$ [$i\geq 2$] of 
Eq.(\ref{Others}) verify the relation: 
\begin{equation}
\sum_{p=1}^\infty\frac{m_\nu^{(p)\ 2}}{m_{KK}^{(p)\ 2}-m^2_{\nu_i}}=-1.
\label{Verify}
\end{equation}
For each solution $m^2_{\nu_i}$ [$i\geq 2$] of Eq.(\ref{Verify}), it is
clear that at least one term of the involved sum must be negative. Strictly 
speaking, for any $i\geq 2$, there exist at least one index $p$ such that
$m_\nu^{(p)\ 2}/(m_{KK}^{(p)\ 2}-m^2_{\nu_i})<0$, or equivalently 
$m_{KK}^{(p)\ 2}<m^2_{\nu_i}$. We thus conclude that all the eigenvalues 
(except the smallest one) are larger than the weakest squared KK mass, 
namely,
\begin{equation}
m^2_{\nu_i}>m_{KK}^{(1)\ 2}
~~~~~~~[i=2,3,4,\dots].
\label{ResultII}
\end{equation}

\renewcommand{\thesubsection}{B.\arabic{subsection}}
\renewcommand{\theequation}{B.\arabic{equation}}
\setcounter{subsection}{0}
\setcounter{equation}{0}

\section{Neutrino mixing angle}
\label{ApMix}

In this appendix, we derive the expression of the neutrino mixing angle 
$\theta_\nu$, defined in Eq.(\ref{DEF}), as function of the elements
entering neutrino mass matrix ${\cal M}$ ({\it c.f.} Eq.(\ref{MassMatrix}))
within the $''+-+''$ scenario.
\\ The definition (\ref{DEF}) of neutrino mixing angle $\theta_\nu$ involves 
the unitary matrix $U$. The definition (\ref{UmatDEF}) of $U$ can be expressed
in a more explicit way as,
\begin{equation}
\psi^{\nu}_{Ln}=\sum_{i=1}^\infty U_{ni} \ \psi^{phys}_{Li}
~~~~~~~[n=0,1,2,\dots], 
\label{UmatDEFbis}
\end{equation}
where the index $i=1,2,3,\dots$ corresponds to the index of mass eigenstates 
$\nu_i$ (see Section \ref{Measure}) and to the index of physical neutrino masses
$m_{\nu_i}$ (see foot-note \ref{index}).
The $i$-th vector $U_{ni}$ represents the eigenstate associated to the eigenvalue
$m^2_{\nu_i}$ of ${\cal M}{\cal M}^\dagger$, and, it is clear from Eq.(\ref{DIAGO}) 
that for each $i$ value (recall that $i=1,2,3,\dots$) one has,
\begin{equation}
\sum_{n=0}^\infty
( {\cal M}{\cal M}^\dagger-m^2_{\nu_i} {\bf 1} )_{mn} 
U_{ni}=0
~~~~~~~[m=0,1,2,\dots]. 
\label{EachEigen}
\end{equation}
After replacing ${\cal M}$ by its expression (\ref{MassMatrix}), in Eq.(\ref{EachEigen}), 
we obtain the following system of equations,
\begin{equation}
\bigg ( \sum_{p=0}^\infty m_\nu^{(p)\ 2}-m^2_{\nu_i} \bigg ) U_{0i}
+ \sum_{n=1}^\infty m_\nu^{(n)} m_{KK}^{(n)} U_{ni}=0,
\label{systemI}
\end{equation}
\begin{equation}
m_\nu^{(m)} m_{KK}^{(m)} \ U_{0i} + (m_{KK}^{(m)\ 2}-m^2_{\nu_i}) U_{mi}=0
~~~~~~~[m=1,2,3,\dots].
\label{systemII}
\end{equation}
The normalization condition for the $i$-th vector $U_{ni}$, namely 
$\sum_{n=0}^\infty U^2_{ni}=1$ [$i=1,2,3,\dots$], leads, together
with Eq.(\ref{systemII}), to the following analytical expression for $U^2_{0i}$,
\begin{equation}
U^2_{0i}=\bigg [ 1+\sum_{m=1}^\infty 
\frac{m_\nu^{(m)\ 2} \ m_{KK}^{(m)\ 2}}{(m_{KK}^{(m)\ 2}-m^2_{\nu_i})^2} \bigg ]^{-1}.
\label{U0I}
\end{equation}
Therefore, a good approximation of the squared matrix element $U^2_{01}$ associated
to the smallest squared neutrino mass eigenvalue $m^2_{\nu_1}$, which verifies 
Eq.(\ref{APPROX}), is,
\begin{equation}
U^2_{01} \simeq \bigg [ 1+\sum_{m=1}^\infty 
\bigg ( \frac{m_\nu^{(m)}}{m_{KK}^{(m)}} \bigg )^2 \ \bigg ]^{-1}.
\label{U01}
\end{equation}
Finally, Eq.(\ref{U01}) and Eq.(\ref{DEF}) allow to obtain the wanted expression
for the neutrino mixing angle $\theta_\nu$: 
\begin{equation}
\tan^2\theta_\nu \simeq \sum_{m=1}^\infty 
\bigg ( \frac{m_\nu^{(m)}}{m_{KK}^{(m)}} \bigg )^2.
\label{wanted}
\end{equation}

\end{document}